\documentclass[a4paper,11pt]{article}
\pdfoutput=1
\usepackage{jheppub} 

\usepackage{amsthm}
\usepackage{amssymb}
\usepackage{amsmath} 
\usepackage{color}
\usepackage{array} 
\usepackage{graphicx}

\usepackage{hyperref}

\newcommand{\be}{\begin{equation}}
\newcommand{\ee}{\end{equation}}

\begin{document}
\begin{titlepage}

\begin{center}


 {\LARGE\bfseries  New gravitational solutions via a \\
  \vskip 3mm 
 Riemann-Hilbert approach }
  \\[10mm]

\textbf{G.L.~Cardoso$^1$, J.C.  Serra$^2$}

\vskip 6mm
{\em  $^1$Center for Mathematical Analysis, Geometry and Dynamical Systems,\\
  Department of Mathematics, 
  Instituto Superior T\'ecnico,\\ Universidade de Lisboa,
  Av. Rovisco Pais, 1049-001 Lisboa, Portugal}\\[4mm]
  
{\em $^2$
 Instituto Superior T\'ecnico,\\ Universidade de Lisboa,
  Av. Rovisco Pais, 1049-001 Lisboa, Portugal}
  \\[4mm]
  
{\tt gcardoso@math.tecnico.ulisboa.pt}
{\tt
joao.leandro.camara.serra@tecnico.ulisboa.pt}

\end{center}

\vskip .2in
\begin{center} {\bf ABSTRACT } \end{center}
\begin{quotation}\noindent 

\noindent
We consider the Riemann-Hilbert factorization approach to solving 
the field equations of 
dimensionally reduced gravity theories.
First we prove that functions belonging to a certain class possess a canonical factorization
due to properties of the underlying spectral curve. Then we use this result, together with appropriate 
matricial decompositions, to study
the canonical factorization of non-meromorphic monodromy matrices that describe
deformations of seed monodromy matrices associated with known solutions. This results in 
new solutions, with unusual features, to the 
field equations.

\vskip 2cm

\end{quotation}
\vfill
\today
\end{titlepage}

\tableofcontents

\section{Introduction}

The Riemann-Hilbert factorization approach to solving the field equations of gravity theories is remarkable in that
it allows to study the subspace of solutions to the field equations that only depend
on two space-time coordinates, in terms of the canonical factorization of a so-called monodromy
matrix into matrix factors $M_{\pm}$ with  prescribed analiticity properties in a complex
variable $\tau$ \cite{Breitenlohner:1986um}.  Thus, instead of directly solving non-linear PDE's in two variables,
the problem of solving the field equations is
mapped to a canonical Riemann-Hilbert factorization problem in one complex variable $\tau$.
Let us describe this approach.

We consider the dimensional reduction of gravity theories (without a cosmological
constant) to two spatial dimensions.
The 
resulting two-dimensional effective action describes a scalar non-linear sigma model coupled to gravity.
The equations of motion for the scalar fields 
can be recast in terms of an auxiliary linear system, called 
the Breitenlohner-Maison linear system \cite{Breitenlohner:1986um}, which depends on an additional parameter,
the spectral parameter $\tau \in \mathbb{C}$. The solvability condition for the linear system
yields the equations of motion for the scalar fields \cite{Breitenlohner:1986um,Schwarz:1995af,Lu:2007jc}, provided $\tau$ is taken to be a position dependent
spectral parameter that satisfies the relation 
\be
\omega = v + \tfrac12 \rho \frac{(1-\tau^2)}{\tau} \;,
\label{spectralc}
\ee
which defines an algebraic curve.
Here $\omega \in \mathbb{C}$, while $(\rho,v) \in \mathbb{R}^2$ denote two 
space-like coordinates, called Weyl coordinates.
Locally, the relation \eqref{spectralc} can be inverted to yield $\tau=\tau(\omega, \rho, v)$, so that $\tau$
is a function of 
the Weyl coordinates that parametrize the two-dimensional space
and of a complex parameter $\omega$. While here we consider the case when $(\rho,v)$ are space-like coordinates,
the case of one time-like and one space-like coordinate can be treated in a similar manner 
\cite{Schwarz:1995af,Lu:2007jc}.

Given a solution to the linear system, one can associate to it a so-called monodromy matrix ${\cal M} (\omega)$,
whose entries only depend on the parameter $\omega$.  Conversely, given a candidate 
monodromy matrix ${\cal M} (\omega)$,
one may ask if, for $\omega$ given by \eqref{spectralc},
 it possesses a so-called canonical factorization, as follows.  
 One considers a closed single contour $\Gamma$ in the complex
  $\tau$-plane, which we take to be the unit circle centered around the origin, i.e. $\Gamma = \{ \tau \in \mathbb{C} : |\tau| =1 \}$. This contour divides the complex plane $\mathbb{C}$ into two regions, 
  namely the interior
and the exterior region of the unit circle $\Gamma$. Then, given a matrix ${\cal M} (\omega (\tau))$, with an inverse
${\cal M}^{-1} (\omega (\tau))$,  
such that both are continuous on $\Gamma$, one seeks a decomposition
  ${\cal M} (\omega(\tau)) = M_-(\tau) M_+ (\tau)$, valid on $\Gamma$, where $M_{\pm}$
have to satisfy certain analiticity and boundedness conditions, to be reviewed in section \ref{sec:review}. In particular,
$M_-$ is analytic in the exterior region, while $M_+$ is analytic in the interior region.

If ${\cal M} (\omega)$ possesses a canonical factorization, then the latter is unique, up to multiplication by a constant
matrix. This freedom can be fixed by imposing a normalization condition on one of the factors, say on $M_+$.
Then, it can be shown \cite{Camara:2017hez} that the factor $M_-$ yields
a solution to the equations of motion for the scalar fields in two dimensions, and consequently of the
gravitational field equations.  The factor $M_+$, on the other hand, yields a solution
to the linear system.

The question of the existence of a canonical factorization for  ${\cal M}(\omega)$ is an example of a matricial 
Riemann-Hilbert factorization problem. In the Riemann-Hilbert approach to gravity, rather than solving the non-linear
field equations directly, one is instructed to perform the canonical factorization of a monodromy matrix 
${\cal M}(\omega)$ in order to obtain a solution to the field equations.  
To be able to obtain explicit solutions to the field equations, the canonical factorization must be performed explicitly.  
While this is, in general, well understood for scalar functions \cite{MP,cc}, the situation changes
dramatically in the case of matrix functions. For the latter, it is not known in general whether
such a factorization exists;
and in case it exists, there are no general methods available for obtaining it explicitly. Thus, 
different methods have to be developed on a case by case basis for different classes of matrices.
The monodromy matrices considered recently in the literature have either simple poles 
\cite{Maison:1988zx,Nicolai:1991tt,Katsimpouri:2012ky,Katsimpouri:2013wka,Katsimpouri:2014ara,Chakrabarty:2014ora}
or double and simple poles \cite{Camara:2017hez} in the $\omega$-plane. 
In \cite{Camara:2017hez} the corresponding matricial Riemann-Hilbert
factorization problem was converted into a vectorial Riemann-Hilbert factorization problem, which was subsequently
solved using a generalization of Liouville's theorem. 
The explicit factorization method that 
was applied to monodromy matrices with single or double poles in \cite{Camara:2017hez} 
is of 
greater generality, but it can be applied only to rational monodromy matrices; the factorization of
non-rational monodromy matrices requires developing other factorization methods.

Which monodromy matrices should one then pick as a starting point?  One strategy for choosing ${\cal M} (\omega)$
consists in starting from the monodromy matrix associated to a known solution to the field equations,
and then deforming this monodromy matrix to obtain a different matrix to be factorized.  This is the
strategy that we will adopt in this paper.  In doing so, we first study 
monodromy matrices
of the form ${\cal M} (\omega) = f(\omega) \, \tilde{\cal M} (\omega)$, where $f(\omega)$ is a scalar function.
Here we obtain our first result, as follows.  Take the unit circle contour $\Gamma$ in the $\tau$-plane,
and consider its image in the $\omega$-plane by means of the algebraic curve relation \eqref{spectralc}.
We will denote the resulting closed contour in the $\omega$-plane by $\Gamma_{\omega}$. 
As we will show in section \ref{sec:review}, any rational function $f$ in $\omega$ 
that has no zeroes and no poles on $\Gamma$
has a canonical factorization in $\tau \in \Gamma$.  Moreover, any function $f(\omega)$ that is continuous 
and non-vanishing on $\Gamma_{\omega}$ also
has a canonical factorization.  This surprising result is quite unexpected in view of the
existing results on the factorization of scalar functions \cite{MP,cc} (which is not, in general,
canonical), and turns out to be
a consequence of the Stone-Weierstrass theorem when combined
with properties of the algebraic curve relation \eqref{spectralc}.

Armed with this result, we turn to the study of solutions
of two four-dimensional gravity theories  in section \ref{newsol}, namely  
Einstein gravity and Einstein-Maxwell-dilaton theory.
The latter is obtained by performing a Kaluza-Klein reduction of five-dimensional Einstein gravity.
We begin with the study of solutions of 
 Einstein-Maxwell-dilaton theory.
As a starting point we pick the solution that describes an $AdS_2 \times S^2$ space-time, supported by electric/magnetic
charges $(Q, P)$ and a constant dilaton scalar field.  The associated monodromy matrix was given in \cite{Camara:2017hez},
and its entries contain double and simples poles at $\omega =0$. We then perform a two-parameter deformation of
this monodromy matrix. The resulting monodromy matrix is of the form ${\cal M} (\omega) = f(\omega) \, \tilde{\cal M} (\omega)$, where $\tilde{\cal M} (\omega)$ is a matrix with entries that are rational functions with
double or simple poles, while $f(\omega)$
is given by the third root of a rational function $g(\omega)$. Since $g(\omega)$ possesses a canonical factorization
in view of the theorem mentioned above, also $f(\omega)$ will possess a canonical factorization, provided we pick the branch
cuts of the roots appropriately. Then, using the explicit factorization method presented in \cite{Camara:2017hez} and
alluded to above, we perform the explicit canonical factorization of $\tilde{\cal M} (\omega)$. The resulting
space-time solution has unusual features. It describes a stationary solution
that is supported by a non-constant
oscillatory dilaton field. It interpolates between an asymptotic space-time with a NUT parameter $J$  and a Killing horizon.
The near-horizon geometry is not $AdS_2\times S^2$,
but nevertheless has a deep throat, and as a result the dilaton field exhibits the behaviour of a static attractor
\cite{Ferrara:1996dd,Ferrara:1996um}: due to the deep throat in the geometry, the dilaton field flows to a constant value that is entirely
specified by the electric/magnetic charges $(Q, P)$.  This unusual stationary solution is complicated: it is
given in terms of a power series in $J$, and it results
from the particular deformation of the monodromy matrix that we chose.
To obtain this solution by
directly solving the field equations is likely not to be straightforward.

Next, we turn to four-dimensional Einstein gravity and study 
deformations of the monodromy matrix associated to the Schwarzschild solution.
The monodromy matrices that arise in four-dimensional Einstein gravity are
 $2 \times 2$ symmetric matrices with entries determined in terms of three functions, which we denote
by $a, b$ and $R^2$, see \eqref{Mmon}. These functions are continuous on $\Gamma_{\omega}$, with
$b^2 R^2 -a^2 =1$ on $\Gamma_{\omega}$. We show that 
any such matrix can be decomposed as in \eqref{MSD} into matrix factors whose entries only
depend on the 
combinations $a \pm b \, R$ and on $R$. To proceed, we have to make a choice for the functions
$R$.  We take $R$ to be a rational function of $\omega$,
with no zeroes and no poles on $\Gamma_{\omega}$, and similarly for its inverse $R^{-1}$.
Note that the Schwarzschild
solution is captured by this class of functions, and this motivates the choice of this class.
With this choice, the combinations $a \pm b \, R$ 
possess a 
canonical factorization 
in view of the theorem that we prove in section \ref{sec:review}
and that we mentioned above. For this choice of functions $R$, we obtain a class
of  monodromy matrices whose 
canonical factorization can be performed
regardless of the type of isolated singularities that the 
combinations $a \pm b \, R$ may have.
We illustrate this by choosing combinations $a \pm b \, R$  that 
have an essential singularity at $\omega =0$.
The resulting monodromy matrix describes
a deformation of the monodromy matrix associated with the Schwarzschild solution.
We show that the presence of this essential singularity does
not pose any problem for explicitly performing the canonical factorization of the deformed
matrix.  This exemplifies that 
 it is possible to perform canonical factorizations of monodromy
matrices that have more complicated singularities than the rational ones 
considered in the recent literature.

Some of the conclusions that we draw from these explicit factorizations are as follows.
First, 
factorization transforms innocently looking deformations of 
monodromy matrices into  
highly non-trivial deformations of space-time solutions.  Second, one may 
ask whether continuity in the deformation parameters is preserved by the factorization.
We have analysed this question for one of the deformation parameters in the example that we studied in the context of
four-dimensional Einstein-Maxwell-dilaton theory, and we find that this is indeed the case.
Third, the deformed space-time solutions that result by using the Riemann-Hilbert factorization
approach may be  difficult to obtain by direct means, i.e. by directly solving the four-dimensional field equations. This constitutes one of the advantages
of the factorization approach. And finally, due to the aforementioned theorem that we 
prove in section \ref{sec:review} and to an appropriate decomposition of the monodromy matrix, 
we find that the difficulties in factorizing
monodromy matrices with singularities
that are not just poles can be overcome.
This in
turn raises a question, which we will not address here: is there a correspondence
between the type of singularities in the $\omega$-plane and properties of the associated space-time
solution?

\section{The Breitenlohner-Maison linear system and canonical factorization \label{sec:review}}

We consider the dimensional reduction of four-dimensional gravity theories (at the two-derivative level,
and without a cosmological constant) down to two dimensions 
\cite{Breitenlohner:1986um}. The resulting theory in two dimensions can be brought to the 
form of a scalar non-linear  sigma-model coupled to gravity. We take the sigma-model 
target space to be a symmetric space $G/H$. 
In performing the reduction, we first reduce 
to three dimensions over a time-like isometry direction. We denote the associated three-dimensional
line element by
\be
ds_3^2 = e^{\psi} \left(d \rho^2 + dv^2 \right) + \rho^2 \, d \phi^2 \;.
\label{met3d}
\ee
Here, $\psi$ is a function of the coordinates $(\rho,v)$, which are called Weyl coordinates.
Throughout this paper, we take $\rho > 0$.
Subsequently, we reduce to two dimensions along the space-like isometry direction $\phi$.

The resulting equations of motion in two dimensions take the form \cite{Breitenlohner:1986um}
\be
d \left( \rho \star A \right) = 0 \;,
\label{eomA}
\ee
where $d$ denotes the exterior derivative, and the matrix one-form  
$A = A_{\rho} \, d \rho + A_v \, d v$ equals 
\be 
A = M^{-1} \, d M \;.
\ee
Here, $M(\rho,v)$ denotes the representative of the symmetric space $G/H$, 
and it satisfies $M = {M}^{\natural}$, where
the operation $\natural$ denotes a 'generalized transposition' that acts
anti-homomorphically on matrices \cite{Lu:2007jc}. The operation
$\star$
denotes the Hodge dual in two dimensions. 
The warp factor $\psi$ in \eqref{met3d} is then obtained by integrating
 \cite{Schwarz:1995af,Lu:2007jc}
\begin{eqnarray}
\partial_{\rho} \psi &=& \tfrac14 \rho \,  {\rm Tr} \left( A^2_{\rho} - A^2_v \right) \;, \nonumber\\
\partial_{v} \psi &=& \tfrac12 \rho \,  {\rm Tr} \left( A_{\rho}  A_v \right) \;.
\label{psi2}
\end{eqnarray}

The equations of motion \eqref{eomA} for $M$ can be reformulated in terms of an auxiliary
linear system, the so-called Breitenlohner-Maison linear system\footnote{In this paper, we work
with the Breitenlohner-Maison linear system. There exists another linear system, the Belinski-Zakharov
linear system, that is often used to address the construction of solitonic solutions.  We refer to 
\cite{Figueras:2009mc}
and references therein for a discussion of the relation between these two linear systems.}, 
whose solvability implies \eqref{eomA}. This linear system reads \cite{Schwarz:1995af,Lu:2007jc}
\be
\tau \, (d + A ) X = \star d X \;.
\label{lax}
\ee
 It depends on a complex parameter $\tau$, called the spectral parameter. To ensure
 that the solvability of \eqref{lax} implies the equations of motion  \eqref{eomA}, $\tau$
 has to satisfy the algebraic curve relation
 \be
 \omega = v + \frac{\rho}{2 \tau} \left(1 - \tau^2 \right) \,,
 \label{omtau}
 \ee
 where $\omega \in \mathbb{C}$. With $\rho >0$, this results in
 \be
 \tau (\omega, \rho, v) = \frac{1}{\rho} \left( v - \omega \pm \sqrt{\rho^2 + (v - \omega)^2} \right) \;\;\;,\;\;\;
 \omega \in \mathbb{C} \;.
 \label{tauom}
 \ee

Let us assume that there exists a pair $(A,X)$ (with $A = M^{-1} dM$) that solves 
\eqref{lax} such that $X$ and its inverse $X^{-1}$ are analytic in $\tau$, for $\tau$
in the interior of the unit circle $\Gamma$ (centered around the origin in the $\tau$-plane), 
with continuous boundary valued funtions
on $\Gamma$.
Then, $A = M^{-1} dM$
is a solution to the equations of motion \eqref{eomA},
and
one
can assign to the pair $(M, X)$ a so-called monodromy matrix ${\cal M} (\omega)$ that satisfies
${\cal M} (\omega) = {\cal M}^{\natural} (\omega)$
and possesses a canonical
factorization, 
\be
{\cal M} (\omega(\tau, \rho, v)) = M_-(\tau, \rho, v) \, M_+ (\tau, \rho, v) \;\;, \;\;\; \tau \in \Gamma \;.
\label{RHpro}
\ee
Here, on the left hand side, $\omega$ is viewed as a function of $\tau \in \mathbb{C}$ using the relation 
\eqref{omtau}, and both ${\cal M}$ and its inverse are continuous on $\Gamma$.
The factorization \eqref{RHpro} is valid in the $\tau$-plane, with respect to the unit circle $\Gamma$
centered around $\tau =0$.
The factors $M_{\pm}$ are such that: $M_+$ and its
inverse $M_+^{-1}$ are analytic and bounded in the interior of the unit circle $\Gamma$,
while  $M_-$ and its
inverse $M_-^{-1}$ are analytic and bounded in the exterior of the unit circle $\Gamma$.
Also, $M_+$ is normalised  to $M_+ (\tau =0) = \mathbb{I}$, rendering the factorization unique. We refer to \cite{Camara:2017hez} for a
comprehensive review of the conditions for the existence of a canonical factorization.

Conversely, consider a monodromy matrix ${\cal M}(\omega)$ that satisfies ${\cal M} = {\cal M}^{\natural} $ and possesses
a canonical factorization \eqref{RHpro} (with $\omega$ given as in \eqref{omtau})
satisfying 
$M_+ (\tau =0) = \mathbb{I}$.
Then, it can be shown that
$M(\rho, v)=M_-(\tau = \infty,
\rho,v)$ is a solution to the equations of motion \eqref{eomA}, and $M_+(\tau, \rho,v)$ is a solution
to the linear system \eqref{lax} \cite{Camara:2017hez}. 
Thus, in this approach,
to obtain explicit solutions to the field equations, we first pick a monodromy matrix 
${\cal M} (\omega)$
that possesses a canonical factorization, then perform the factorization explicitly to extract
the factors $M_{\pm}$, to obtain $M(\rho, v)=M_-(\tau = \infty,
\rho,v)$, which encodes the space-time solution.
Note that the condition \eqref{RHpro}, or equivalently the jump condition ${\cal M} \, M_+^{-1} = M_-$ on $\Gamma$, defines
a matrix Riemann-Hilbert problem.

Necessary and sufficient conditions for the existence of a canonical factorization 
${\cal M} = M_- \, M_+$ were summarized in \cite{Camara:2017hez},
where a method for obtaining explicit factorizations was also described. This method, based on solving a vectorial
Riemann-Hilbert problem ${\cal M}  \, \phi_+ = \phi_-$ by means of Liouville's theorem (or a generalization thereof), is the one that we will 
follow throughout. Here, the $\phi_+$ denote the columns of $M_+^{-1}$, while the $\phi_-$
denote the columns of $M_-$.

The monodromy matrices that we will factorize will be of the type ${\cal M}(\omega) = f(\omega) \, {\tilde {\cal M}}
(\omega)$, where $f(\omega)$ denotes a function, and ${\tilde {\cal M}}$ a matrix.  Since we are interested
in monodromy matrices that have a canonical factorization, i.e. a factorization of the form \eqref{RHpro}
with $M_{\pm}$ satisfying the analyticity and boundedness properties described in the text below \eqref{RHpro},
which class of functions $f(\omega)$ should we consider
to ensure that $f(\omega)$ has a canonical factorization (i.e. $f(\omega) = f_+ (\tau) f_-(\tau)$
with $f_{\pm}$ satisfying the analyticity and boundedness properties described in the text below \eqref{RHpro})?  
Here we obtain the following surprising result: \\

{\sl Theorem: } Any function $f(\omega)$ that is continuous and non-vanishing on $\Gamma_{\omega}$ has a canonical
factorization. Here, $\Gamma_{\omega}$ denotes the image of the curve $\Gamma$ (the unit circle in the $\tau$-plane
centered at $\tau =0$)
under \eqref{omtau}.

\begin{proof} The proof uses the Stone-Weierstrass theorem and the algebraic curve relation \eqref{omtau}.

The Stone-Weierstrass theorem states that if $K$ is a compact subset of $\mathbb{C}$, then every continuous,
complex-valued function $f$ on $K$ can be uniformly approximated by polynomials $P_n$ in $\omega$ and $\bar \omega$, i.e.
\be
\sup_{\omega \in K} | f(\omega) - P_n(\omega, \bar \omega)| \rightarrow 0 \;\;\;,\;\; n \rightarrow \infty \;\;\;,\;\;\; n \in \mathbb{N}
\;.
\ee
Here we take $K = \Gamma_{\omega}$, where $\Gamma_{\omega}$ denotes the image of the curve 
$\Gamma$ under \eqref{omtau}.
Then, using \eqref{omtau}, we obtain the relation  ${\bar \omega} = - \omega + 2v$, which holds
 for any $\omega \in \Gamma_{\omega}$.
Thus, viewed as functions of $\omega \in \Gamma_{\omega}$,  the ${\tilde P}_n(\omega) = 
P_n(\omega, \bar \omega)$
are polynomials in $\omega$, while viewed as functions of $\tau \in \Gamma$, 
the ${\tilde P}_n(\omega (\tau)) $ are rational functions in $\tau$. We will show momentarily that any rational function
in $\omega$, when seen as a function of $\tau \in \Gamma$ via composition with $\omega (\tau)$,
 has a canonical factorization in $\tau \in \Gamma$, so long as it has no zeroes and no poles on $\Gamma$. Then, 
since $ \sup_{\tau \in \Gamma} | f(\omega(\tau) ) - {\tilde P}_n(\omega (\tau)) | = 
 \sup_{\omega \in K} | f(\omega) - {\tilde P}_n(\omega) | $, $f(\omega(\tau))$ is uniformly approximated by rational functions on $\Gamma$ that
possess a canonical factorization. Therefore, $f$ also has a canonical factorization 
(see, for instance, section 5 of the review paper \cite{cc} and references therein).

The same conclusion is reached by looking at the index of the function $f(\omega(\tau)) $.
This index is defined as follows \cite{MP}. We consider a function $f(\omega)$ that is continuous and non-vanishing 
on $\Gamma_{\omega}$, and hence also
continuous and non-vanishing on $\Gamma$. 
Let us 
represent the image of the function $f(\omega(\tau))$ in a complex plane, which we will call the 
$F$-plane. We denote the image of $\Gamma$ under $f(\omega(\tau))$ by
$\Gamma_f$. Then, $\Gamma_f$ is a closed contour in the $F$-plane
that does not pass through the origin of the $F$-plane. The index of $f$ is then defined to be
the winding number of the closed contour $\Gamma_f$ around the origin of the $F$-plane.
If this winding number is zero, then $f(\omega(\tau)) $ possesses a canonical factorization 
by a known theorem \cite{MP},
which states that any continuous function on $\Gamma$ with zero index admits a canonical factorization.
We proceed to show that the winding number is zero.

The relation \eqref{omtau} associates two values of $\tau$, given by \eqref{tauom},
to any $\omega$.  Denoting these two values by $\tau_1$ and $\tau_2$, we have
$\tau_1 \tau_2 = -1$. This means that there are two values of $\tau \in \Gamma$
that correspond to the same $\omega \in \Gamma_{\omega}$, namely $\tau_1 \in \Gamma$
and $- {\bar \tau}_1 \in \Gamma$.  Now consider going around $\Gamma$ once, counterclockwise, starting
at $\tau = -i$. In doing so,  let us denote the directed curve starting at $\tau = -i$ and ending at $\tau = i$ by $\Gamma_1$, while the directed curve starting at $\tau = i$ and ending at $\tau =-i$ will
be denoted by $\Gamma_2$. If we denote the image of $\Gamma_1$ under  \eqref{tauom}
by $\gamma_{\omega} \subset \Gamma_{\omega}$, then the image of $\Gamma_2$ under  \eqref{tauom}
is $- \gamma_{\omega}$.  Therefore, if we go around $\Gamma$ once in a counterclockwise fashion,
the closed curve $\Gamma_{\omega}$ that is travelled in the $\omega$-plane is $\gamma_{\omega} - \gamma_{\omega}$.  
Hence, if $f(\omega)$ is continuous and non-vanishing 
on $\Gamma_{\omega}$, the resulting contour $\Gamma_f$ has zero winding number with respect to the origin of
the $F$-plane, and we conclude that $f(\omega(\tau))$ has a canonical factorization.

Finally, let us show that any rational function in $\omega$ has a canonical 
factorization in $\tau \in \Gamma$, 
so long as it has no zeroes and no poles on $\Gamma$. We begin by writing
\eqref{omtau} in the form
\begin{equation}
\omega - \omega_0 = -\frac{\rho(\tau - \tau_0^+)(\tau - \tau_0^-)}{2\tau} \;,
\end{equation}
where $\tau_0^+$, $\tau_0^-$ are the two values of $\tau$ corresponding to $\omega = \omega_0$. We assume that $\tau_0^{\pm}$ do not lie on the unit circle $\Gamma$ in the $\tau$-plane, and we take
 $\tau_0^+$ to lie inside and $\tau_0^-$ to lie outside the unit disc.  Now consider a 
 rational function $f(\omega)$ with $A_1$
zeroes and $A_2$ poles (counting multiplicities). When viewed as a function of $\tau$ using
 \eqref{omtau}, 
$f(\omega(\tau))$
takes the form
\begin{equation}
f(\omega(\tau)) = \big (-\frac{\rho}{2\tau}\big )^{A_1 - A_2} \; \frac{\prod\limits_{i=1}^{A_1}(\tau - \tau_i^+)}
{\prod\limits_{j=1}^{A_2}(\tau - \tau_j^+)}\frac{\prod\limits_{i=1}^{A_1}(\tau - \tau_i^-)}{\prod\limits_{j=1}^{A_2}(\tau - \tau_j^-)} \;,
\end{equation}
up to an overall normalization constant.  Here we assume that none of the zeroes 
and poles of $f(\omega(\tau)) $ lie on the unit circle $\Gamma$, so that
$f(\omega(\tau)) $ and its inverse are continuous on $\Gamma$. 
Then, 
$\tau_i^+$ and $\tau_i^-$ are the two values of $\tau$ corresponding to the zero $\omega = \omega_i$, with $|\tau_i^+| < 1$ and $|\tau_i^-| > 1$,
and similarly for the two values $\tau_j^+$ and $\tau_j^-$ associated with the pole $\omega = \omega_j$. The function $f(\omega(\tau))$ possesses a canonical factorization,
$ f = f_- \, f_+$, if $f_-$ and its inverse are analytic in the exterior of the unit disc and bounded at $\tau = \infty$, and if $f_+$ and its inverse are analytic and bounded in the interior of the unit disc .  This is indeed the case, as can be verified by taking
\be
f_- = 
\big (-\frac{\rho}{2\tau}\big )^{A_1 - A_2} \; \frac{\prod\limits_{i=1}^{A_1}(\tau - \tau_i^+)}
{\prod\limits_{j=1}^{A_2}(\tau - \tau_j^+)} \;\;\;,\;\;\;
f_+ = \frac{\prod\limits_{i=1}^{A_1}(\tau - \tau_i^-)}{\prod\limits_{j=1}^{A_2}(\tau - \tau_j^-)} \;.
\ee
\end{proof}
Note that, while it is true that any rational function in $\tau$ has a factorization, the latter is in general not canonical (see, for instance, \cite{cc} and references therein). However, in our case, due to the spectral curve relation between $\omega$ and $\tau$, any rational function in $\omega$
has a canonical factorization in $\tau$,
provided that it has no zeroes nor poles on the unit circle $\Gamma$.

In particular, 
if a rational function $f(\omega(\tau))$ 
has a canonical factorization, $f = f_- \, f_+$,
 then its $n$-th root also has one. 
This is a consequence of the following textbook theorem \cite{Ahl}: if $g(z)$ is analytic
and non-vanishing in a simply connected region $\Omega$, then it is possible to define
single-valued analytic branches of $\sqrt[\leftroot{+3}\uproot{3}n]{g(z)}$ in $\Omega$.  
Setting $g = f_+$, we obtain part of the assertion.  For the factor $f_-$, we use the Schwarz reflection principle and consider
${\tilde f}_+ = {\bar f}_-$, defined in the interior region of $\Gamma$ by ${\tilde f}_+ (z) = \overline{f _- ( 1/{\bar z} )}$ and on $\Gamma$ by ${\tilde f}_+ (z) = \overline{f _- ( z)}$, which is analytic
(extended to $z=0$) and non-vanishing in the unit disk. Hence, we can apply the theorem mentioned above to ${\tilde f}_+$ to define an analytic branch of $\sqrt[\leftroot{+3}\uproot{3}n]{{\tilde f}_+(z)}$.
Then, applying  the Schwarz reflection principle to the latter, we obtain
 $\overline{\left( \sqrt[\leftroot{+3}\uproot{3}n]{ \overline{ {f}_-(z)  }} \right)}$, which defines
an analytic branch of $\sqrt[\leftroot{+3}\uproot{3}n]{f_-(z)}$, and the assertion follows.

Given a solution $M(\rho, v)$ of the two-dimensional equations of motion \eqref{eomA},
it may be useful to have a rule that assigns a 
candidate monodromy matrix ${\cal M} (\omega)$ to it.  One such rule is the 
so-called substitution rule given in \cite{Camara:2017hez}, which consists
in the following. Assuming that the limit $\lim_{\rho \rightarrow 0^+} M(\rho, v)$ exists,
the candidate monodromy matrix is obtained by substituting $v$ by $\omega$ in this expression,
i.e.
\begin{equation}
{\cal M} (\omega = v) = \lim_{\rho \rightarrow 0^+} M(\rho, v) \;.
\end{equation}
Whether, upon performing a canonical factorization, this candidate monodromy matrix really yields
back the solution $M(\rho, v)$ has to be verified case by case.

\section{New solutions by deformation of seed monodromy matrices \label{newsol}}

In this section we construct new solutions to the dimensionally reduced gravitational
field equations by deforming the monodromy matrices associated to known solutions.
The monodromy matrices of the latter will be called seed monodromy matrices.
The deformed monodromy matrices we consider fall into the class of monodromy matrices
to which the theorem given in the previous section applies, and hence they 
possess a canonical factorization, which we carry out explicitly.  We do this in the context
of two gravitational theories, namely four-dimensional Einstein-Maxwell-dilaton theory (obtained
by Kaluza-Klein reduction of five-dimensional Einstein gravity)
and four-dimensional Einstein gravity theory.

We begin by considering the Einstein-Maxwell-dilaton theory.

\subsection{Deformed monodromy matrices in dimensionally reduced
Einstein-Maxwell-dilaton theory \label{H1H2ex}}
The field equations of 
the four-dimensional Einstein-Maxwell-dilaton theory, that is obtained 
by Kaluza-Klein reducing five-dimensional Einstein gravity, admits extremal black hole solutions \cite{Rasheed:1995zv,Matos:1996km,Larsen:1999pp,Astefanesei:2006dd,Emparan:2007en,Chow:2014cca}.
These solutions, which 
may be static 
or rotating, 
are supported by a scalar field $e^{- 2 \Phi}$ (the dilaton field)
and by an electric charge $Q$ and a magnetic charge
$P$. We will take $Q>0,  P >0$ throughout. An example of a static solution is the extremal Reissner-Nordstr\"om black hole solution, which is supported by a constant dilaton field $e^{- 2 \Phi} = Q/P$,
and which
interpolates
between flat space-time and 
an $AdS_2 \times S^2$ space-time. In adapted coordinates, the four-dimensional line element of the latter reads 
\begin{equation}
ds_4^2 = - \frac{r^2}{{Q} {P}} \, dt^2+ {Q} {P} \, \frac{dr^2}{r^2} + {Q} { P}
\left( d \theta^2 + \sin^2 \theta \, d \phi^2 \right) \;.
\label{adss2}
\end{equation}
In these coordinates, the extremal Reissner-Nordstr\"om solution
 interpolates between a flat space-time metric 
at $r = \infty$ and a near-horizon metric \eqref{adss2} at $r =0$.

The $AdS_2 \times S^2$ space-time, described by \eqref{adss2} and supported by a constant
dilaton field $e^{- 2 \Phi} = Q/P$, is by itself a solution to the field equations of the Einstein-Maxwell-dilaton theory.
As shown in \cite{Camara:2017hez}, upon dimensional reduction to two dimensions, this solution can 
be associated with the following monodromy matrix,
\begin{eqnarray}
{\cal M}_{\rm seed} (\omega ) =
  \begin{pmatrix}
A/{\omega}^2   & \quad B/{\omega}  &\quad  C \\
- B/{\omega}   &\quad D
& \quad 0 \\
C &\quad  0 & \quad 0 
\end{pmatrix} \;\;\;,\;\;\; \det {\cal M} = 1 \;,
\label{monomatrixMseed}
\end{eqnarray}
where
\begin{equation}
A = {P}^{4/3} \,  {Q}^{2/3} \;\;\;,\;\;\; B = \sqrt{2} \, {P}^{1/3} \, {Q}^{2/3} \;\;\;,\;\;\;
 C = - \left( \frac{P}{Q} \right)^{1/3} \;\;\;,\;\;\; D = - \left( \frac{Q}{P} \right)^{2/3} \;.
\end{equation}
Note that the entries in \eqref{monomatrixMseed} are rational functions in the variable
$\omega \in \mathbb{C}$, and that $A, B >0$, while $C, D <0$. Also observe that 
$2 AD + B^2 =0$ and $- C^2 D = 1$. Note that ${\cal M}_{\rm seed}$ satisfies ${\cal M}^{\natural}_{\rm seed} = 
{\cal M}_{\rm seed}$, 
where the anti-homomorphism $\natural$ denotes a 'generalized transposition' that is not simply transposition, see 
\cite{Chakrabarty:2014ora} for details.

The monodromy matrix \eqref{monomatrixMseed}, which we call the seed monodromy matrix,
can be deformed in different ways. If the resulting monodromy matrix has a canonical
factorization, then, as explained in section \ref{sec:review}, its factorization will yield a solution
to the field equations 
of the theory.  In \cite{Camara:2017hez}, a specific
deformation of \eqref{monomatrixMseed}
was considered that gave rise to the extremal Reissner-Nordst\"om 
black hole solution mentioned above.
This deformation of the monodromy matrix 
was implemented by the transformation
\begin{equation}
{\cal M}(\omega) = g^{\natural} \, {\cal M}_{\rm seed} \, g \;\;\;,\;\;\; g = e^N \;,
\label{sol-gen}
\end{equation}
with $N$ a constant nilpotent matrix 
which resulted in \cite{Camara:2017hez}
\begin{eqnarray}
{\cal M} (\omega ) =
  \begin{pmatrix}
A/{\omega}^2 + A_1/\omega + A_2  &\qquad B/{\omega} + B_2 &\qquad  C \\
- B/{\omega} - B_2  &\qquad D
& \qquad 0 \\
C &\qquad  0 & \qquad 0 
\end{pmatrix} \;\;\;,\;\;\; \det {\cal M} = 1 \;,
\label{monomatrixMg}
\end{eqnarray}
with non-vanishing constants $A_1, A_2, B_2$ satisfying
\begin{equation}
B_2^2 = - 2 A_2 \, D \;\;\;,\;\;\; A_1 = - (B_2 \, B)/D \;.
\end{equation}
Note that $A_2 > 0$. Taking $B_2 >0$, this yields $A_1 = 2 \sqrt{A \, A_2}$, where we used $B = \sqrt{-2 A\, D}$.
Then, using $A \sqrt{-D} = P Q $ and defining $h = \sqrt{A_2 \sqrt{-D}} > 0$, we obtain
\begin{eqnarray}
{\cal M} (\omega ) =
  \begin{pmatrix}
H^2(\omega)/{\sqrt{-D}}&\qquad  \sqrt{2} \, \sqrt{\sqrt{-D}} \, H(\omega)
&\qquad  -1/{\sqrt{-D}} \\
- \sqrt{2} \, \sqrt{\sqrt{-D}} \, H(\omega) &\qquad D
& \qquad 0 \\
-1/{\sqrt{-D}} &\qquad  0 & \qquad 0 
\end{pmatrix} \;,
\label{monomatrixMg2}
\end{eqnarray}
where
\be
H(\omega) = h + \frac{\sqrt{P Q} }{\omega} \;.
\ee
The associated space-time solution describes an  extremal Reissner-Nordst\"om 
black hole with line element
\begin{equation}
ds_4^2 = - \frac{1}{H^2(r)} \, dt^2  + H^2(r) \Big( dr^2 + r^2 
\left( d \theta^2 + \sin^2 \theta \, d \phi^2 \right) \Big)\;,
\label{staticinter}
\end{equation}
supported by a constant scalar field $e^{-2 \Phi}
= - D^{3/2} =Q/P$.

In the following, we consider a different deformation of \eqref{monomatrixMseed}. We replace
$(Q,P)$ in \eqref{monomatrixMseed} by
\be
Q \rightarrow Q + h_1 \, \omega \;\;\;,\;\;\; P \rightarrow P + h_2 \, \omega \;,
\ee
where we view $(h_1, h_2) \in \mathbb{R}^2 $ as deformation parameters. We restrict to $h_1>0, h_2>0$ throughout.
We then obtain
the monodromy matrix ${\cal M}={\cal M}^{\natural}$, 
\begin{eqnarray}
{\cal M}(\omega)  =
\left( \frac{H_2}{H_1} \right)^{1/3} 
  \begin{pmatrix}
H_1 H_2   &\quad  \sqrt{2} H_1  &\quad  - 1 \\
- \sqrt{2} H_1  &\quad  - H_1/H_2 
& \quad  0 \\
- 1 &\quad  0 & \quad 0 
\end{pmatrix} \;\;\;,\;\;\; \det {\cal M} = 1 \;,
\label{monomatrixMHH}
\end{eqnarray}
where
\be
H_1 (\omega)  = h_1 + \frac{ Q}{\omega} \;\;\;,\;\;\;  H_2 (\omega) =  h_2 + \frac{ P}{\omega} \;.
\ee
This deformed monodromy matrix is of the type ${\cal M}(\omega) = f(\omega) \, {\tilde {\cal M}}(\omega)$, where 
${\tilde {\cal M}}(\omega)$ is a matrix with rational entries, and $f(\omega)$ is the third root of a rational function.
Thus, according to the theorem of section \ref{sec:review}, ${\cal M}(\omega)$ possesses a canonical factorization,
which below we carry out explicitly.

We note that the deformed monodromy matrix \eqref{monomatrixMHH} is not of the form 
 $g^{\natural}  \, {\cal M}_{\rm seed} \, g$, with $g$ given by
 $g (\omega) = e^{N(\omega)}$, where
  $N (\omega)$ is a nilpotent,  possibly $\omega$-dependent lower triangular matrix. As shown 
  in \cite{Camara:2017hez}, 
  such a  transformation would result
  in a matrix of the form \eqref{monomatrixMg}, with $\omega$-dependent coefficients $A_1, A_2, B_2$.
  This does not reproduce \eqref{monomatrixMHH}. Whether there exists a $g(\omega)$ such that the
  transformation $g^{\natural} (\omega) \, {\cal M}_{\rm seed} (\omega) \, g (\omega) $ reproduces
  \eqref{monomatrixMHH} is an open question that we will not address here.

Since the deformed monodromy matrix \eqref{monomatrixMHH} is not of the form 
 $g^{\natural}  \, {\cal M}_{\rm seed} \, g$, with $g$ a constant matrix, 
  the resulting space-time
 solution lies outside of the class of solutions considered in \cite{Rasheed:1995zv,Larsen:1999pp,Chow:2014cca}.
The resulting space-time solution, which has a Killing horizon, will be stationary whenever the combination 
$J = h_1 P - h_2 Q$ is non-vanishing. 
The near-horizon
solution, however, 
will exhibit the behaviour of a static attractor.
We proceed to explain this.

Let us compare 
\eqref{monomatrixMHH} with \eqref{monomatrixMg2}. They will agree when
\be
h_1 P = h_2 Q \;\;\;,\;\; h = \sqrt{h_1 h_2} \;.
\ee
The condition $h = \sqrt{h_1 h_2}$ is a normalization condition that we will pick in the following.
Therefore, only when $h_1 P = h_2 Q $ does the
associated space-time solution describe the static solution 
\eqref{staticinter} that is 
supported by a constant dilaton field. When $h_1 P \neq h_2 Q $, the interpolating space-time solution 
will not any longer remain static, and the dilaton field will cease to be constant.   The solution will acquire
a complicated dependence on the angular coordinate $\theta$, as will be shown below.
The four-dimensional line element associated with this solution takes the form
\begin{equation}
ds^2_4 = - e^{-\phi_2}\, \left(dt + {\cal A}_{\phi} \, d\phi \right)^2 + e^{\phi_2} \left(e^{\psi} \, (dr^2 + r^2 \,
d \theta^2) + r^2 \, \sin^2 \theta \, d\phi^2 \right) \;,
\label{linegenral}
\end{equation}
where the functions $\phi_2, \psi$ and the one-form ${\cal A } = {\cal A}_{\phi} \, d \phi$ depend on the coordinates $(r, \theta)$.
When $ h_1 P = h_2 Q $, $\phi_2$ becomes a function of $r$ only, while $\psi =0$ and ${\cal A}_{\phi} =0$, 
and the line element reduces to the one in \eqref{staticinter}.  However, when  $ h_1 P \neq  h_2 Q $, $\psi$ 
and ${\cal A}$ are not any longer zero.  They become non-trivial functions of $(r, \theta)$ that
are given in terms of series expansions in the parameter ${\tilde J} = J/(h_1 h_2)$, where
\be
{\tilde J } = {\tilde P} - {\tilde Q} \;,
\ee
with ${\tilde Q} = Q/h_1 \;,\; {\tilde P} = P/h_2 $. We recall that ${\tilde Q} >0, {\tilde P}>0$.

 To assess the impact of a non-vanishing $\tilde J$ on the solution, one may consider
taking $\tilde J$ to be small and working to first order in $\tilde J$.  In doing so, we find the following.
Asymptotically, as $r \rightarrow \infty$, we have $e^{\phi_2} \rightarrow h^2 \,, \, e^{\psi} \rightarrow 1 $, while ${\cal A}_{\phi} = - J
\, \cos \theta $. Thus, the asymptotic geometry is stationary, with a NUT parameter $J$.  
On the other hand, when approaching the Killing horizon at $r = 0$, the dilaton field tends to the constant value
$e^{2 \Phi} = P/Q$, while 
$\psi$ tends to zero and $e^{\phi_2}$ behaves as $QP/r^2$.
However, ${\cal A}_{\phi}$ tends to $J \, \cos \theta (1 - \cos \theta)$, and therefore the 
near-horizon geometry does not have the isometries of $AdS_2 \times S^1$
or $AdS_2 \times S^2$ \cite{Astefanesei:2006dd}. Nevertheless, as $r \rightarrow 0$, the dilaton exhibits the behaviour of a static attractor \cite{Ferrara:1996dd,Ferrara:1996um}:
due to the deep throat in the geometry, the dilaton field flows to a constant value that is entirely specified
by the electric/magnetic charges.

Now we proceed with the canonical factorization of ${\cal M}(\omega)$ given in \eqref{monomatrixMHH}.
To perform the factorization explicitly, we will use the vectorial factorization method mentioned in section 
\ref{sec:review}. 
Inspection of \eqref{monomatrixMHH} shows that the monodromy matrix has poles in the 
$\omega$-plane located at $\omega =0, - {\tilde P}, - {\tilde Q}$. Using the spectral curve relation \eqref{tauom},
these values correspond to the following values on the $\tau$-plane,
\begin{eqnarray}
\tau_0^{\pm} &=& \frac{1}{\rho}\Big (v \pm \sqrt{\rho^2 + v^2}\Big ) \;, \nonumber\\
\tau_{\tilde P}^{\pm} &=& \frac{1}{\rho} \left(v + {\tilde P} \pm \sqrt{\rho^2 + (v+{\tilde P})^2} \right) \;,
\nonumber\\
\tau_{\tilde Q}^{\pm} &=& \frac{1}{\rho} \left(v + {\tilde Q}
\pm \sqrt{\rho^2 + (v+ {\tilde Q})^2} \right) \;.
\label{tvval}
\end{eqnarray}
We recall $\rho > 0$, and we take $v \in \mathbb{R} \backslash 
\{ 0, - \tilde{P}, - \tilde{Q} \}$, so that the values \eqref{tvval}
are never on the unit circle in the $\tau$-plane.  Note that $\tau^+_0 \tau ^-_0 = -1$, 
and similarly for the
other two pairs in 
\eqref{tvval}. Thus, given any of the pairs in \eqref{tvval}, one of the values lies inside,
while the other lies outside of the unit circle. The values inside of the unit circle will be denoted by $\tau^+$,
while the values outside of the unit circle will be denoted by $\tau^ -$. 
Depending on the region in the $(\rho, v)$-plane,
$\tau^+$ may either correspond to the $+$ branch or the $-$ branch in \eqref{tvval}.
The coordinates $(\rho, v)$ are related to the coordinates
$(r, \theta)$ by
 \begin{equation}
 \rho = r \, \sin \theta \;\;\;,\;\;\; v = r \, \cos \, \theta \;,
 \label{rtrv}
 \end{equation}
where $r >0$ and $ 0 < \theta < \pi$.

The monodromy matrix \eqref{monomatrixMHH} is the product of a matrix ${\tilde {\cal M}}(\omega)$ 
with a scalar factor
$f(\omega)$. Then, the canonical factorization of ${\cal M}(\omega)=M_- (\tau) M_+(\tau)$ is obtained\footnote{We suppress the dependency of $M_{\pm}$ on $(\rho, v)$ for notational simplicity.} 
by performing the canonical factorization
of $f(\omega) = f_-(\tau) \, f_+ (\tau) $ and of ${\tilde {\cal M}}(\omega)
=
{\tilde M}_- (\tau) {\tilde M}_+(\tau)$, so that
$M_-(\tau) = f_-(\tau) \, {\tilde M}_- (\tau)$ and $M_+(\tau) = f_+(\tau) \, {\tilde M}_+ (\tau)$.
Note that  $M_+(\tau)$ has to satisfy the normalization condition  $M_+(\tau=0) = \mathbb{I}$.

The scalar factor $f(\omega) = \Big(H_2(\omega)/H_1(\omega)\Big )^{1/3} $
has the canonical factorization 
\be
f_-(\tau) =  \left(\frac{h_2}{h_1}\right)^{1/3} \Big(
\frac{\tau_{\tilde P}^{-}}{\tau_{\tilde Q}^{-}}
\Big )^{1/3} 
\Big (
\frac{\tau - \tau_{\tilde P}^{+}}{\tau - \tau_{\tilde Q}^{+}}
\Big )^{1/3}
\;\;\;,\;\;\; 
f_+ (\tau) =  \Big(
\frac{\tau_{\tilde P}^{+}}{\tau_{\tilde Q}^{+}}
\Big )^{1/3} 
\Big(
\frac{\tau - \tau_{\tilde P}^{-}}{\tau - \tau_{\tilde Q}^{-}}
\Big )^{1/3} \;,
\ee
where we imposed the normalization $f_+ (\tau =0)=1$.
Here, the branch cuts are the line segments connecting $\tau_{\tilde P}^+$ with $\tau_{\tilde Q}^+$ 
and $\tau_{\tilde P}^-$ with $\tau_{\tilde Q}^-$, and we take $1^{1/3} = 1$. We note
\be
 f_- (\tau=\infty) = 
  \left(\frac{h_2}{h_1}\right)^{1/3} \Big(
\frac{\tau_{\tilde P}^{-}}{\tau_{\tilde Q}^{-}}
\Big )^{1/3} \;.
\label{pppm}
\ee

Next, we factorize 
\begin{eqnarray}
\tilde{\cal M}(\omega) =
  \begin{pmatrix}
H_1 H_2   &\quad  \sqrt{2} H_1  &\quad  - 1 \\
- \sqrt{2} H_1  &\quad  - H_1/H_2 
& \quad  0 \\
- 1 &\quad  0 & \quad 0 
\end{pmatrix} \;,
\end{eqnarray}
by using the vectorial factorization method mentioned in section \ref{sec:review}, which is
set up in the form
\be
{\tilde {\cal M}} \, {\tilde M}_+^{-1} = {\tilde M}_- \;,
\label{matrixtilH}
\ee
and which consists in considering the columns $\phi_{+}$ of ${\tilde M}_+^{-1}$ and the columns $\phi_{-}$ of ${\tilde M}_-$
and solving the associated vectorial factorization problem,
i.e.
\begin{equation}
 {\tilde {\cal M}} \, \phi_{+}  = \phi_{-} \;, 
 \label{vecfact}
\end{equation}
column by column. In doing so, we use a generalized version of Liouville's theorem, and impose the
normalization condition ${\tilde M}_+ (\tau =0) = \mathbb{I}$. We refer to appendix 
\ref{explfact} for the details of the factorization.

Having obtained the factorization ${\cal M}(\omega)=M_- (\tau) \, M_+(\tau)$, we extract the matrix $M(\rho,v)$
that contains the space-time information, 
\begin{eqnarray}
M(\rho,v) = M_- (\tau = \infty) =   g
  \begin{pmatrix}
m_1  &\quad  m_2 &\quad  - 1 \\
- m_2  &\quad  m_3
& \quad  0 \\
- 1 &\quad  0 & \quad 0 
\end{pmatrix} \;,
\label{matr}
\end{eqnarray}
where
\begin{eqnarray}
g &=&  f_- (\tau=\infty) =  \Big(\frac{h_2}{h_1}\Big)^{1/3} \Big(
\frac{\tau_{\tilde P}^{-}}{\tau_{\tilde Q}^{-}}
\Big )^{1/3}  \;, \nonumber\\
m_1 &=& h_1 h_2  \left( 1 - \frac{2 {\tilde Q}}{\rho (\tau_0^+ - \tau_0^-)}  \right)
  \left( 1 - \frac{2 {\tilde P}}{\rho (\tau_0^+ - \tau_0^-)}  \right) 
   - 2 h_1 h_2  \, \frac{(\tau_{\tilde Q}^+ - \tau_{\tilde P}^+ ) (\tau_0^+ - \tau_{\tilde P}^-)
(\tau_0^+ - \tau_{\tilde Q}^+)}{\tau_{\tilde Q}^+ (\tau_0^+ - \tau_0^- )^2} \;,
    \nonumber\\
m_2 &=& 
\sqrt 2 \, h_1  \left( 1 - \frac{2 {\tilde Q}}{\rho (\tau_0^+ - \tau_0^-)}  \right)
- \sqrt{2} \, h_1  \, \frac{(\tau_{\tilde Q}^+ - \tau_{\tilde P}^+ ) 
(\tau_0^+ - \tau_{\tilde Q}^+)}{\tau_{\tilde Q}^+ (\tau_0^+ - \tau_0^- )} \;,
\nonumber\\
m_3 &=& - \frac{h_1}{h_2} \, \frac{\tau_{\tilde Q}^-}{\tau_{\tilde P}^-} = 
 - \frac{h_1}{h_2}  \left( 1 +  \frac{\tau_{\tilde Q}^- - \tau_{\tilde P}^- }{\tau_{\tilde P}^-} \right) \;.
 \label{m123}
\end{eqnarray}
We note the relation
\begin{equation}
g^3 \, m_3 = -1 \;.
\end{equation}

Next, we
relate the line element \eqref{linegenral}
to  $M(\rho,v)$. We begin with the 
warp factor $\psi$, which, when viewed as a function of $(\rho, v)$,  is obtained by integrating
\eqref{psi2},
\begin{eqnarray}
\partial_{\rho} \psi &=& \tfrac14 \rho \,  g^4 \, \left[ \Big( \partial_{\rho} (g \, m_3) \Big)^2 
-  \Big( \partial_{v} (g \, m_3) \Big)^2 \right]
\;, \nonumber\\
\partial_{v} \psi &=& \tfrac12  \, \rho \, g^4 \, \partial_{\rho} (g \, m_3)
 \, \partial_v (g \, m_3)   \;.
 \label{psieq}
\end{eqnarray}
Taking mixed derivatives of these equations, it can be verified that they are consistent, and hence
\eqref{psieq} can be integrated.  Below we will solve \eqref{psieq} in terms of a formal 
series expansion in powers of ${\tilde J}$,
\begin{equation}
\psi (\rho, v) = \sum_{n=2}^{\infty} \psi_n(\rho, v)  \, {\tilde J}^n \;,
\label{psiser}
\end{equation}
up to a constant term which we set to zero by imposing the normalization condition $\psi =0$ when ${\tilde J} =0$.  Note that there is no linear term in  ${\tilde J}$.
The $\psi_n$ will be determined by expanding the right hand side of \eqref{psieq}
in powers of ${\tilde J}$, and integrating these equations order by order in ${\tilde J}$.
We will then verify that when ${\tilde J}$ is small, it suffices to keep the first few terms in the series \eqref{psiser} to obtain
an expression for $\psi$ that is excellent agreement with the exact solution of \eqref{psieq}.

To relate the warp factor $e^{-\phi_2}$, the one-form ${\cal A}$ and the dilaton field $e^{- 2 \Phi}$
to the data contained in $M(\rho,v)$, we use 
the following parametrization of $M(\rho,v)$ as a coset representative of $SL(3, \mathbb{R})/SO(2,1)$ 
\cite{Chakrabarty:2014ora},
\begin{eqnarray}
M(\rho,v) =  \begin{pmatrix}
e^{2\Sigma_1}   & \quad  e^{2\Sigma_1} \, \chi_2  & \quad e^{2\Sigma_1} \, \chi_3 \\
- e^{2\Sigma_1} \, \chi_2  & \quad - e^{2\Sigma_1} \, \chi_2^2 +  e^{2\Sigma_2}
& \quad  - e^{2\Sigma_1} \, \chi_2 \, \chi_3 +  e^{2\Sigma_2} \, \chi_1\\
e^{2\Sigma_1} \, \chi_3 & \quad e^{2\Sigma_1} \, \chi_2 \, \chi_3 -  e^{2\Sigma_2} \, \chi_1
 & \quad - e^{2\Sigma_2} \, \chi_1^2 +  e^{2\Sigma_1} \, \chi_3^2 + e^{2\Sigma_3}
\end{pmatrix} \;,
\label{cosetrepM}
\end{eqnarray}
where
\begin{eqnarray}
\Sigma_1 &=& \frac12 \left( \frac{1}{\sqrt{3}} \phi_1 + \phi_2 \right) \;, \nonumber\\
\Sigma_2 &=& - \frac{1}{\sqrt{3}} \phi_1 \;, \nonumber\\
\Sigma_3 &=& \frac12 \left( \frac{1}{\sqrt{3}} \phi_1 - \phi_2 \right) \;, 
\label{sigphi}
\end{eqnarray}
which satisfies $\Sigma_1 + \Sigma_2 + \Sigma_3 =0$. 
Then
\begin{equation}
\phi_2 = 2 \Sigma_1 + \Sigma_2 \;,
\label{fss}
\end{equation}
while the two-form ${\cal F} = d \cal A$ is determined by
\begin{equation}
- e^{- 2 \phi_2} *{\cal F} = d \chi_3 - \chi_1 \, d \chi_2 \;.
\label{acc}
\end{equation}
Here, the dual $*$ is with respect to the three-dimensional metric \eqref{met3d}.
The dilaton field is given by $e^{- 2 \Phi} = e^{- \sqrt{3} \, \phi_1/2}= e^{3 \Sigma_2/2}$.

Comparing \eqref{cosetrepM} with \eqref{matr}, we infer
\begin{eqnarray}
e^{2 \Sigma_1} &=& m_1 \, g \;\;\;,\;\;\; e^{2 \Sigma_2}  = (m_3 + (m_2)^2/m_1) \, g
\;\;\;,\;\;\; e^{- 2 \Sigma_3}  = (m_1 \, m_3 + (m_2)^2) \, g^2 \;, \nonumber\\
\chi_1 &=& - m_2/(m_1 \, m_3 + (m_2)^2) \;\;\;,\;\;\; \chi_2 = m_2/m_1 \;\;\;,\;\;\; \chi_3 = -1/m_1 \;.
\end{eqnarray}
This results in the expressions
\begin{eqnarray}
e^{2 \phi_2} &=&  g^3 \, m_1 \, (m_1 \, m_3 + (m_2)^2) \;, \nonumber\\
e^{- 2 \Phi} &=& g^{3/2} \, \left(m_3 + \frac{(m_2)^2}{m_1} \right)^{3/2} \;,
\nonumber\\
d \chi_3 - \chi_1 \, d\chi_2 &=& \frac{1}{m_1 (m_1 \, m_3 + (m_2)^2)} \, \left(m_3 \, dm_1+ m_2 \, d m_2\right) \;, \nonumber\\
* {\cal F} &=& dm_1 + \frac{m_2}{m_3} \, dm_2 \;.
\label{warps}
\end{eqnarray}
It can be verified that $d * \left( dm_1 + \frac{m_2}{m_3} \, dm_2 \right) =0$, so that $d {\cal F} =0$,
and hence ${\cal F} = d {\cal A}$, locally.

Finally, we recall that 
the solution carries electric-magnetic charges $(Q,P)$, and hence 
is supported by an electric-magnetic field, which is described by a one-form $A^0$
that can be read off by performing the dimensional reduction of Einstein gravity in five dimensions to four dimensions
\cite{Chakrabarty:2014ora}, and
given by
\begin{equation}
A^0 = \chi_1 \, dt + A_{\phi} \, d \phi \;,
\end{equation}
with $A_{\phi}$ determined by
\begin{eqnarray}
\partial_{\rho} A_{\phi} &=& - e^{2(\Sigma_1 - \Sigma_2)} \, \rho \, \partial_{v} \chi_2
+ {\cal A}_{\phi} \, \partial_{\rho} \chi_1 \;, \nonumber\\
\partial_{v} A_{\phi} &=& e^{2(\Sigma_1 - \Sigma_2)} \, \rho \, \partial_{\rho} \chi_2
+ {\cal A}_{\phi} \, \partial_{v} \chi_1 \;.
\label{emfield1}
\end{eqnarray}

Now we turn to the interpretation of the space-time solution described by line element \eqref{linegenral}, by 
\eqref{warps}, \eqref{emfield1} and by \eqref{psieq}. To this end, and for concreteness,
we focus on the region $\rho >0, v>0$,
so that also $v + {\tilde Q} >0$, $v + {\tilde P} >0$. Then, 
\begin{eqnarray}
\rho \, \tau_0^+ &=& v - \sqrt{v^2 + \rho^2} \;, \nonumber\\
\rho \,\tau_{\tilde Q}^+ &=& v + {\tilde Q}- \sqrt{(v + {\tilde Q})^2 + \rho^2} \;, \nonumber\\
\rho \, \tau_{\tilde P}^+ &=& v + {\tilde P}- \sqrt{(v + {\tilde P})^2 + \rho^2} \;.
\label{taurhovreg}
\end{eqnarray}
First we consider the series expansion \eqref{psiser}.
As we mentioned below \eqref{psiser}, we determine the explicit form of the $\psi_n$ by expanding
the right hand side of \eqref{psieq}
in powers of ${\tilde J}$, and subsequently  integrating these equations order by order in ${\tilde J}$.
In this way, we find that 
the first terms in this series, which begins with $n=2$, are given by
\begin{eqnarray}
\psi_2 (\rho, v)  &=& - \frac{\rho^2}{18 \,  [ ( {\tilde Q} + v)^2 + \rho^2]^2} \;, \nonumber\\
\psi_3 (\rho, v)   &=& \frac{\rho^2 \, ({\tilde Q} + v)}{9 \, [ ( {\tilde Q} + v)^2 + \rho^2]^3} \;, \nonumber\\
\psi_4 (\rho, v)   &=& \frac{\rho^2 \, [ - 24 \, ({\tilde Q} + v)^2 + 5 \rho^2] }{144 \, [ ( {\tilde Q} + v)^2 + \rho^2]^4} \;.
\end{eqnarray}
We now verify that if we only keep these first terms in the series expansion  \eqref{psiser},
the resulting function $\psi = \sum_{n=2}^{4} \psi_n (\rho, v) \, {\tilde J}^n$ is in excellent agreement with the exact solution of \eqref{psieq}
for small values of ${\tilde J}$. We do this by comparing the expression on the left hand side
of \eqref{psieq}, computed using $\psi = \sum_{n=2}^{4} \psi_n (\rho, v) \, {\tilde J}^n$, with
the exact expression on the right hand side of \eqref{psieq}. This is depicted in Figures 1 and 2
for various ranges of $(\rho,v)$, where
we plotted the difference of the two expressions appearing
in the first equation of \eqref{psieq} for the values ${\tilde Q} = 1, {\tilde J} = 0.001$.

\begin{figure}
  \caption{Behaviour near $\rho=0, v=0$, 
     in the range $0 \leq \rho \leq 0.1 \; , \; 0 \leq v \leq 0.1$.}
  \centering
    \includegraphics[width=0.5\textwidth]{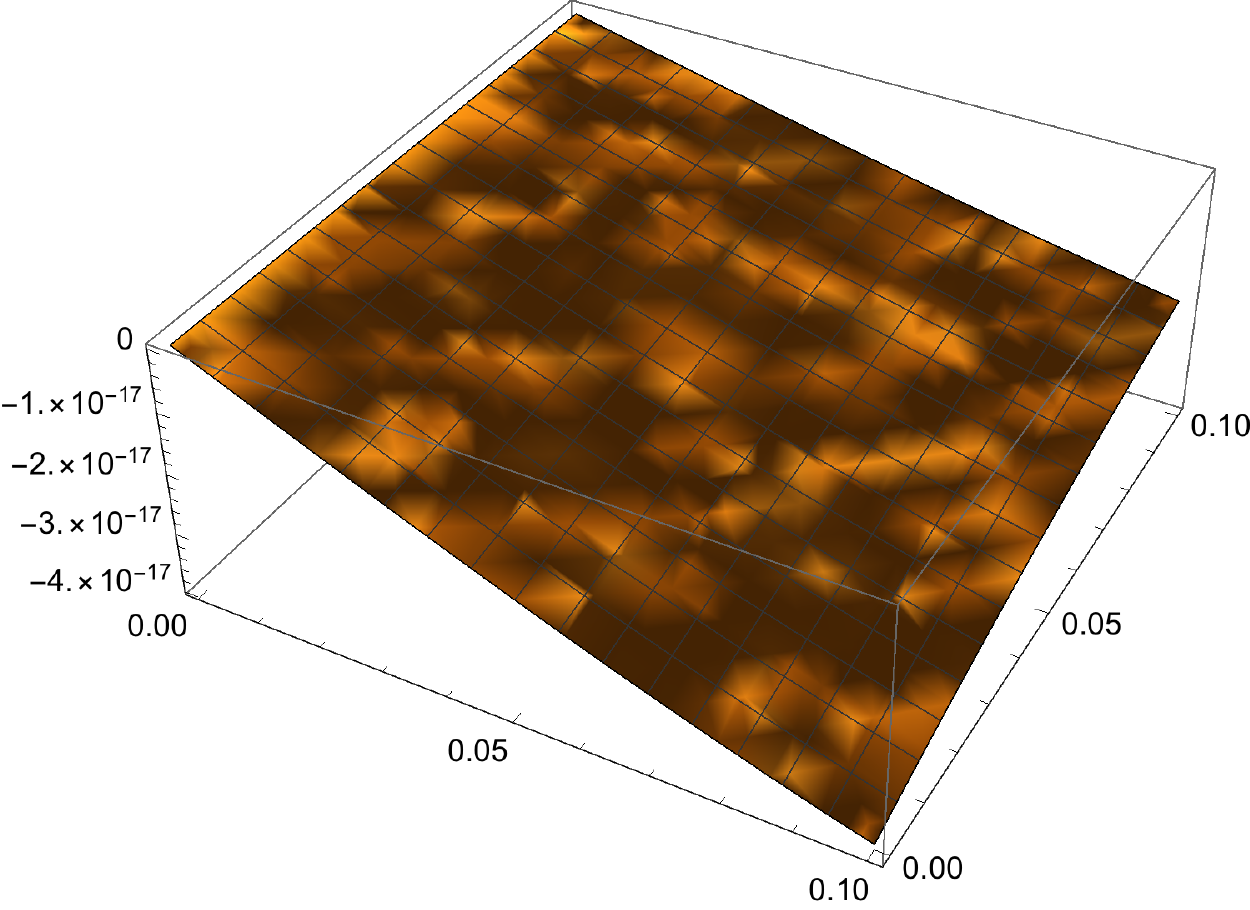}
\end{figure}

\begin{figure}
  \caption{Behaviour in the range $0.1 \leq \rho \leq 10 \; , \; 0.1 \leq v \leq 10$.}
  \centering
    \includegraphics[width=0.5\textwidth]{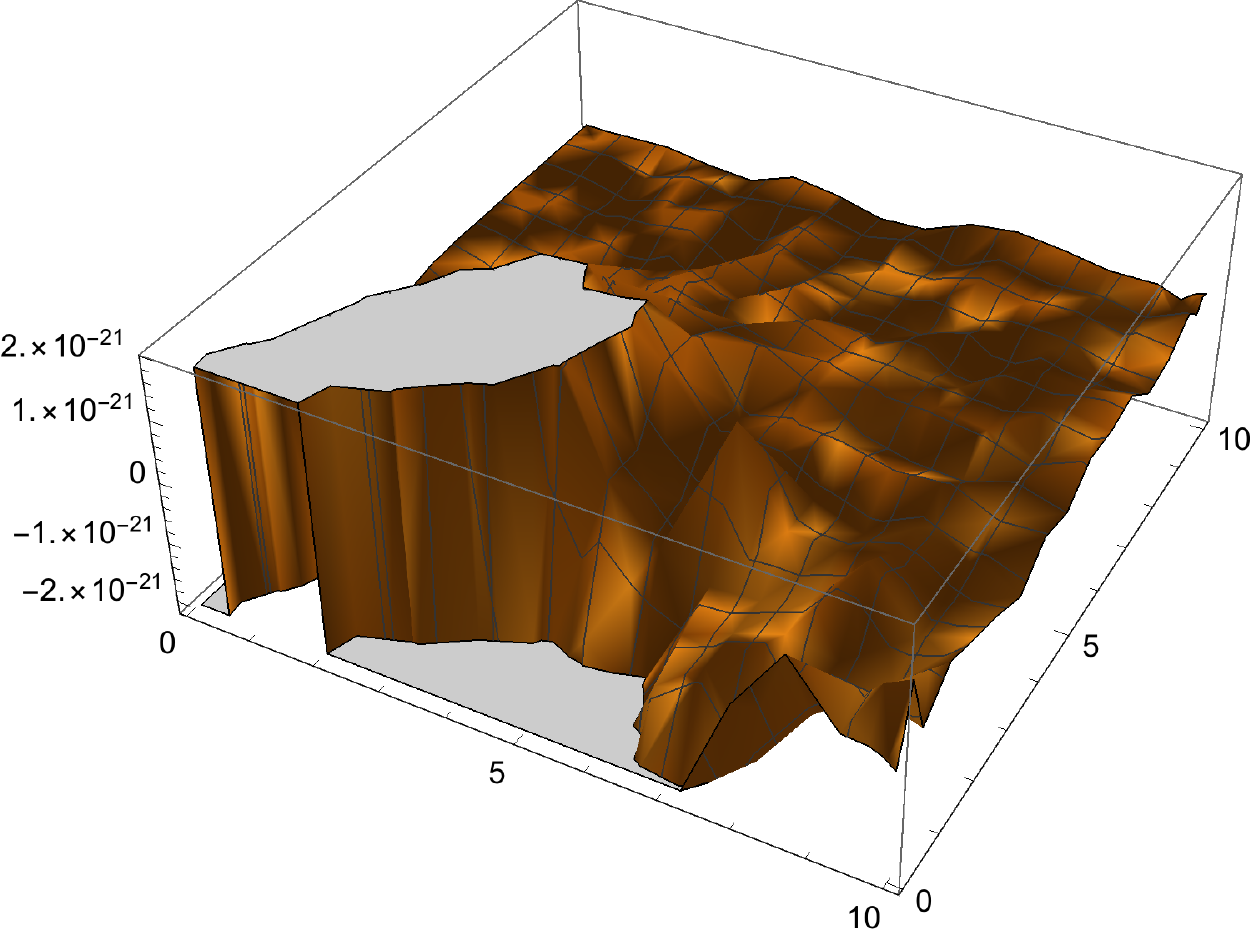}
\end{figure}

In the following, we work at first order in $\tilde J$, in which case $\psi$ may be approximated by 
$\psi =0$.
Then, from \eqref{warps} we infer
\begin{equation}
{\cal A}_{\phi} = - J \, \frac{\left(\rho^2 + v ({\tilde Q} + v) \right) \left(v - \sqrt{\rho^2 + v^2} \right)}
{(\rho^2 + v^2) \sqrt{\rho^2 + ({\tilde Q} + v)^2}}  + {\cal O}({\tilde J}^2 )\;,
\end{equation}
up to a constant, 
while from \eqref{emfield1} we obtain
\begin{equation}
A_{\phi} = \sqrt{2} \, h_2 \left[
\left( {\tilde Q} + \frac{{\tilde J}}{2} \right) \frac{v}{\sqrt{\rho^2 + v^2} }
+  \frac{{\tilde J}}{2} \, \frac{\sqrt{\rho^2 + ({\tilde Q} + v)^2} \left(v - \sqrt{\rho^2 + v^2} \right)}{
\rho^2 + v^2 + {\tilde Q} \sqrt{\rho^2 + v^2} } \right] + {\cal O}({\tilde J}^2 ) \;,
\end{equation}
up to a constant.

To assess the impact of a non-vanishing $\tilde J$ on the solution, we switch to coordinates $(r, \theta)$ given by \eqref{rtrv}. 
Asymptotically, as $r \rightarrow \infty$, we have $e^{- 2 \Phi} \rightarrow h_1/h_2 \;,\;
e^{\phi_2} \rightarrow h_1 h_2 \:,\; e^{\psi} = 1 $, while ${\cal A}_{\phi} = - J
\, \cos \theta $ and $A_{\phi} = \sqrt{2} \, P \, \cos \theta$, up to a constant. Thus, the asymptotic solution 
is stationary, with a NUT parameter $J$,
and is reminiscent of four-dimensional extremal solutions with NUT charge \cite{Bergshoeff:1996gg,Behrndt:1997ny}
that are related to five-dimensional extremal solutions
through the 4d/5d lift \cite{LopesCardoso:2007qid}.

On the other hand, when approaching  the Killing horizon $r = 0$, the dilaton field tends to the constant value
$e^{2 \Phi} \rightarrow {P}/Q $, while 
$e^{\phi_2}$ tends to $Q P/r^2$, $A_{\phi}$ tends to $\sqrt{2} \, P \, \cos \theta$ and
${\cal A}_{\phi}$ tends to $J \, \cos \theta (1 - \cos \theta)$. Hence, the resulting geometry
does not have the isometries of $AdS_2 \times S^2$. 
Nevertheless, the dilaton behaves like a static attractor with charges $(Q,P)$:
the dilaton flows to a value entirely determined in terms of these charges, and the area of
the Killing horizon at $r=0$ (which has zero angular velocity)  equals
$4 \pi \, Q P$, and is thus also determined in terms of the charges. 
The Kretschmann scalar and $R^{\mu \nu} R_{\mu \nu}$ are finite at $r=0$. This static attractor behaviour 
at $r \rightarrow 0$ is then an
unusual feature of this stationary solution.
It can be verified that corrections of order ${\tilde J}^2$ do not affect the attractor behaviour just described.

The behaviour just described may raise the 
following question.  
The canonical factorization of ${\cal M}(\omega)  $
yields factors $M_{\pm} (\tau, \rho, v)$, from which one extracts the space-time information $M(\rho, v)$  by means of
$M(\rho, v) = M_-(\tau = \infty, \rho, v)$.  Now consider the 
coordinates
$(r, \theta)$ given in \eqref{rtrv}.
Since asymptotically, as $r \rightarrow \infty$, the matrix $M(\rho, v)$ tends
to a constant matrix, this may naively suggest that asymptotically, space-time is static and flat.
This naive expectation is, however, incorrect. To obtain the asymptotic form of the space-time line element, one also needs to calculate
the one-form ${\cal A}$ that appears in the line element \eqref{linegenral}. This is done by first calculating 
the field strength ${\cal F} = d {\cal A}$, whose components are 
expressed in terms of derivatives of the entries of $M(\rho, v)$, see \eqref{warps}. 
To first order in ${\tilde J}$, 
${\cal F}$
exhibits an asymptotic  falloff of the form $J/r$. The associated 
potential ${\cal A}_{\phi}$ is thus non-vanishing,
${\cal A}_{\phi} = - J \cos \theta$.

Next, we
verify the validity of the substitution rule. As mentioned in section \ref{sec:review},
this is done by considering $M(\rho, v)$ in the limit $\rho \rightarrow 0^+$,
and subsequently setting $v = \omega$.  If the resulting matrix equals ${\cal M}(\omega)$,
the substitution rule holds in the example under study. To take the limit $\rho \rightarrow 0^+$,
we need to specify a region in parameter space $(\rho, v)$.  We focus again 
on the region $v>0$, for concreteness. We recall that 
${\tilde Q} >0$ ,  ${\tilde P}>0$. Then, using \eqref{taurhovreg},
we obtain in the limit $\rho \rightarrow 0^+$,
\begin{eqnarray}
 \tau_0^+ &=& - \frac{\rho}{2v} \;\;\;,\;\;\;  \tau_0^- = \frac{2v}{\rho} \;,
  \nonumber\\
\tau_{\tilde Q}^+ &=&  - \frac{\rho}{2(v + {\tilde Q})} \;\;\;,\;\;\;  \tau_{\tilde Q}^- = \frac{2(v + {\tilde Q})}{\rho} 
\;, \nonumber\\
\tau_{\tilde P}^+ &=& - \frac{\rho}{2(v + {\tilde P})}  
\;\;\;,\;\;\;  \tau_{\tilde P}^- = \frac{2(v + {\tilde P})}{\rho} 
 \;.
\end{eqnarray}
Using this, we infer
\begin{eqnarray}
\lim_{\rho \rightarrow 0^+} M(\rho, v) = {\cal M}(v) \;,
\end{eqnarray}
thus verifying the validity of the substitution rule in the region  $v>0$.

Finally, we address the following question: the monodromy matrix \eqref{monomatrixMHH}
depends on the parameters $h_1, h_2$. When factorizing, the factors $M_-$ and $M_+$ will
also depend on these parameters. Is this dependence a continuous one? This is not entirely obvious.
Consider, for instance the limit $h_2 \rightarrow 0$: 
 the zero $\omega = - {P}/h_2$ of $H_2$
  in \eqref{monomatrixMHH}
 will move
to infinity.  We will verify that the factorization of  \eqref{monomatrixMHH}
with $h_2 =0$ gives results that coincide with those obtained by factorizing 
 \eqref{monomatrixMHH} with $h_2 \neq 0$ and subsequently taking the limit 
 $h_2 \rightarrow 0$.  

Let us first consider the case when 
$h_1 \rightarrow 0, h_2 \rightarrow 0$.
We consider the region $v>0$, and we recall ${Q}/h_1 > 0, {P}/h_2 > 0$.
In the limit  $h_1 \rightarrow 0, h_2 \rightarrow 0$, we have
\begin{eqnarray}
\tau_{\tilde Q}^+ \rightarrow - \frac{ \rho \, h_1}{2 {Q}} \rightarrow 0 \;\;\;,\;\;\;
\frac{\tau_{\tilde Q}^+}{h_1} \rightarrow - \frac{ \rho }{2 {Q}} \;\;\;,\;\;\;
h_1 \, \tau_{\tilde Q}^- \rightarrow \frac{2 {Q}}{\rho} \;,
\end{eqnarray}
and similarly for ${P}$. We obtain
\begin{eqnarray}
A =  \frac{4 { Q} {P} }{\rho^2  (\tau_0^+ - \tau_0^-)^2}  
      \;\;\;,\;\;\;
B = - 2
 \sqrt 2  \, 
 \, \frac{Q
}{ \rho (\tau_0^+ -  \tau_0^-)  
} 
\;\;\;,\;\;\;
C = - \frac{Q}{P} \;,
\label{hhzero}
\end{eqnarray}
in agreement with \cite{Camara:2017hez}.

Next, we keep $h_1 \neq 0$, and send $h_2 \rightarrow 0$.
We again consider the region $v>0$, with ${Q}/h_1 > 0, {P}/h_2 > 0$.
We obtain
\begin{eqnarray}
A &=&  \left( h_1 - \frac{2 {Q}}{\rho (\tau_0^+ - \tau_0^-)}  \right)
  \left(  - \frac{2 {P}}{\rho (\tau_0^+ - \tau_0^-)}  \right) 
   +4 h_1 {P}   \, \frac{ 
(\tau_0^+ - \tau_{\tilde Q}^+)}{\rho  (\tau_0^+ - \tau_0^- )^2}
    = - 2 h_1 {P}   \, \frac{ 
(\tau_{\tilde Q}^+ - \tau_{\tilde Q}^-)}{\rho  (\tau_0^+ - \tau_0^- )^2} \;, \nonumber\\
B &=& 
 \sqrt 2 h_1 \,
 \, \frac{ \left[  \tau_0^+   -
\tau_{\tilde Q}^- 
 \right] 
}{(\tau_0^+ -  \tau_0^-)  
} \;,
\nonumber\\
C &=& - \frac{h_1}{2 {P}} \, \rho \, \tau_{\tilde Q}^-  \;.
\end{eqnarray}
Comparing with the explicit factorization results that we obtain when factorizing \eqref{monomatrixMHH} with $h_2 =0$,
we find perfect agreement.  
Thus, we have verified that the factorization depends in a continuous manner on the
parameter $h_2$: there is no breakdown when $h_2 =0$ (keeping $h_1 \neq 0$).


\subsection{Deformed monodromy matrices in dimensionally reduced
Einstein gravity  \label{Einsex}}
Next, we consider four-dimensional Einstein gravity reduced to two space-like dimensions.
The associated monodromy matrices are $2 \times 2$ matrices that satisfy
\begin{equation}
{\cal M} = {\cal M}^{\natural} = {\cal M}^T \;\;\;,\;\;\; \det {\cal M} = 1 \;,
\end{equation}
where, is this case, the operation $\natural$ is transposition.
The monodromy matrix is thus a symmetric matrix, which we write in the form
\begin{eqnarray}
{\cal M } = 
\begin{pmatrix}
b & \; a \\
a  & \; b  \, R^2
\end{pmatrix}
\;.
\label{Mmon}
\end{eqnarray}
The monodromy matrix 
depends on three continuous functions  $a, b$ and $R^2$, with
$b^2 R^2 - a^2 =1$ on $\Gamma_{\omega}$.
${\cal M }$ can be decomposed as
\begin{eqnarray}
{\cal M }
=  
\Sigma \, D \, {\Sigma}^{-1} \, J \;,
\label{MSD}
\end{eqnarray}
where
\begin{eqnarray}
\Sigma = 
\begin{pmatrix}
1 & \; 1\\
R & \; -R
\end{pmatrix} \;\;\;,\;\;\;
D = 
\begin{pmatrix}
a+bR & 0\\
0 & a-bR
\end{pmatrix} 
\;\;\;,\;\;\; 
J = 
\begin{pmatrix}
0 & \; 1\\
1 & \; 0
\end{pmatrix} \;.
\end{eqnarray}
Here we assumed $R \neq 0$, so that $\Sigma$ is invertible,
\begin{eqnarray}
\Sigma^{-1} = \frac{1}{2R} \begin{pmatrix}
R & \; 1\\
R & \; -1
\end{pmatrix} \;.
\end{eqnarray}
Note that the case $R \equiv 0$ requires $a$ to be imaginary (since $\det {\cal M}=1$).
We discard this case in the following.

Now let us discuss the canonical factorization of  \eqref{MSD}.
The monodromy matrix \eqref{Mmon} depends on three 
functions  $a, b$ and $R^2$
that, in the decomposition \eqref{MSD}, get assembled into combinations
 $a \pm b \, R$ and $R$. 
To proceed, we have to make a choice for the functions
$R$.  We take $R$ to be a rational function of $\omega$ that is bounded at $\omega = \infty$, 
with no zeroes and no poles on $\Gamma_{\omega}$,
and similarly
for its inverse $R^{-1}$. Note that the Schwarzschild
solution is captured by this class of functions $R$.  For this class of functions $R$, the combinations $a \pm b \, R$ are continuous
functions of $\omega$ and non-vanishing on 
$\Gamma_{\omega}$, and hence possess  a
canonical factorization according to the theorem in section \ref{sec:review}.
Now let us discuss the various factors 
in the decomposition \eqref{MSD}:

\begin{enumerate}

\item 
 
The diagonal matrix $D= {\rm diag} \, (d_1, d_2)$, which is determined in terms of the combinations 
$d_1 = a +b \, R, d_2 = a - b \, R$,
has a canonical factorization, $D = D_- D_+$,
\begin{eqnarray}
D_- = 
\begin{pmatrix}
d_{1-} & 0\\
0 & d_{2-}
\end{pmatrix}
\;\;\;,\;\;\;
D_+ = 
\begin{pmatrix}
d_{1+} & 0\\
0 & d_{2+}
\end{pmatrix} \;.
\end{eqnarray}
Since $D$ is diagonal, this is a scalar factorization problem.

\item The functions $R$ and $R^{-1}$ are rational functions of $\omega$ that  are bounded at $\omega = \infty$. We normalize $R(\infty)=1$. Writing $R(\omega) = r(\omega)/s(\omega)$,
we take $r(\omega)$ and $s(\omega)$ to be both polynomials of degree $n$.
Then, using the algebraic curve 
\eqref{omtau},
the resulting function $R(\omega(\tau))$ and its inverse will be rational functions in $\tau$ 
with $2n$ poles and $2n$ zeroes (counting multiplicities).
For concreteness, we take $R(\omega(\tau))$ to have $2n$ simple poles located at 
$\tau_i^{\pm}$ ($i = 1, \dots, n$). The 
$n$ poles at 
$\tau_i^+$ are located in the interior of the unit disc, and the $n$ poles 
$\tau_i^-$ are located in the exterior of the unit disc in the $\tau$ plane.
Thus, we take $R(\omega(\tau))$ to be
\begin{equation}
R(\omega(\tau)) = \frac{q(\tau)}{ \prod_{i=1}^n(\tau - \tau_i^+)(\tau - \tau_i^-)} \;,
\end{equation}
where $q(\tau)$ denotes a polynomial of degree $2n$ (with $n$ simple zeroes inside
the unit circle, and $n$ simple zeroes outside the unit  circle),
satisfying the normalization condition $q(\tau) = \tau^{2n}$ at $\tau = \infty$, so that $R(\omega(\infty)) =1$.

\end{enumerate}

Next, let us discuss the canonical factorization\footnote{We suppress the dependency of $M_{\pm}$ on $(\rho, v)$ for notational simplicity.} 
of ${\cal M}(\omega)$,
\begin{equation}
{\cal M}(\omega (\tau))  = M_- (\tau) \, M_+ (\tau) \;\;\;,\;\;\; \tau \in \Gamma \;.
\label{mmpm}
\end{equation}
We begin by noting that \eqref{mmpm}
can be written as
\begin{equation}
D_+ \, \Sigma^{-1} \, J \, M_+^{-1} = D_-^{-1} \, \Sigma^{-1} \, M_- \;.
\end{equation}
Denoting
\begin{eqnarray}
M_- = 
\begin{pmatrix}
\phi_{1-}^1 & \phi_{1-}^2\\
\phi_{2-}^1 & \phi_{2-}^2
\end{pmatrix}
\;\;\;,\;\;\;
M_+^{-1} = 
\begin{pmatrix}
\phi_{1+}^1 & \phi_{1+}^2\\
\phi_{2+}^1 & \phi_{2+}^2
\end{pmatrix} \;,
\end{eqnarray}
we obtain
\begin{equation}
\begin{pmatrix}
d_{1 +} & 0\\
0 & d_{2 +}
\end{pmatrix}
\begin{pmatrix}
R & 1\\
R & - 1
\end{pmatrix}
\begin{pmatrix}
\phi_{2+}^1 & \phi_{2+}^2\\
\phi_{1+}^1 & \phi_{1+}^2
\end{pmatrix}
= 
\begin{pmatrix}
1/d_{1 -} & 0\\
0 & 1/d_{2 -}
\end{pmatrix}
\begin{pmatrix}
R & 1\\
R & -1
\end{pmatrix}
\begin{pmatrix}
\phi_{1-}^1 & \phi_{1-}^2\\
\phi_{2-}^1 & \phi_{2-}^2
\end{pmatrix} \;.
\end{equation}
This yields 
the following pair of linear systems ($S_1$ and $S_2$), 
\begin{eqnarray}
&& S_1: \left\{
\begin{array}{rcl}
 R \, d_{1+}\phi_{2+}^1 + d_{1+}\phi_{1+}^1  = \left( 
R \, \phi_{1-}^1 + \phi_{2-}^1 
\right)/d_{1-} \;, \vspace{2mm}
\nonumber\\
R \, d_{2+}\phi_{2+}^1 - d_{2+}\phi_{1+}^1  =  \left( R \, \phi_{1-}^1 - \phi_{2-}^1 
\right)/d_{2-} \;,
\end{array}\right.
\nonumber\\
&& S_2: \left\{
\begin{array}{rcl}
 R \, d_{1+}\phi_{2+}^2 + d_{1+}\phi_{1+}^2  = 
 \left( 
R \, \phi_{1-}^2 + \phi_{2-}^2 \right)/d_{1-} \;, \vspace{2mm}
\\
  R \, d_{2+}\phi_{2+}^2 - d_{2+}\phi_{1+}^2 = 
\left( R\,  \phi_{1-}^2 - \phi_{2-}^2 \right)/d_{2-} \;.
\end{array}\right. 
\end{eqnarray}
Here we have separated the functions that are analytic inside the unit disc in the $\tau$-plane
from those that are analytic in the outside region of the unit disc. The only exception to this is the function $R$.
By assumption, $R(\omega(\tau)) $ is a rational function satisfying $R(\omega(\infty))=1$,
with $2n$ simple zeroes and $2n$ simple poles.
Then,  by means of a generalization of Liouville's theorem (see
 the lemma on page 14 of \cite{Camara:2017hez}), we have
 \begin{eqnarray}
 \label{Spm}
&&S_1: \left\{
\begin{array}{rcl}
 R \, d_{1+}\phi_{2+}^1 + d_{1+}\phi_{1+}^1  =  \left( 
R \, \phi_{1-}^1 + \phi_{2-}^1 
\right)/d_{1-}
= \frac{p_1 (\tau) }{ \prod_{i=1}^n(\tau - \tau_i^+)(\tau - \tau_i^-)} \;, \vspace{2mm}
\nonumber\\
 R \, d_{2+}\phi_{2+}^1 - d_{2+}\phi_{1+}^1  =  \left( R \, \phi_{1-}^1 - \phi_{2-}^1 
\right)/ d_{2-}
= \frac{p_2 (\tau) }{ \prod_{i=1}^n(\tau - \tau_i^+)(\tau - \tau_i^-)} \;,
\end{array}\right. \nonumber\\
&& S_2: \left\{
\begin{array}{rcl}
  R \, d_{1+}\phi_{2+}^2 + d_{1+}\phi_{1+}^2  = 
 \left( 
R \, \phi_{1-}^2 + \phi_{2-}^2 \right)/d_{1-}
= \frac{p_3 (\tau) }{ \prod_{i=1}^n(\tau - \tau_i^+)(\tau - \tau_i^-)} \;, \vspace{2mm}
\nonumber\\
 R \, d_{2+}\phi_{2+}^2 - d_{2+}\phi_{1+}^2  = 
 \left( R\,  \phi_{1-}^2 - \phi_{2-}^2 \right)/d_{2-}
= \frac{p_4 (\tau) }{ \prod_{i=1}^n(\tau - \tau_i^+)(\tau - \tau_i^-)} \;,
\end{array}\right.
\nonumber\\
\end{eqnarray}
 where $p_j(\tau), j=1,2,3,4$ are polynomials of degree $2n$ in $\tau$, to ensure boundedness
 at $\tau = \infty$. 
 They contain a total of $8n + 4$ constants. 
 These constants are determined by
 \begin{enumerate}
 \item imposing the normalization condition $M_+ (\tau =0) = \mathbb{I}$, which yields the 
 four normalization conditions
 \begin{equation}
 \phi_{1+}^1 (\tau =0) = 1 \;\;,\;\; \phi_{2+}^2 (\tau =0) = 1 \;\;,\;\;
  \phi_{2+}^1 (\tau =0) = 0 \;\;,\;\; \phi_{1+}^2 (\tau =0) = 0 \;;
 \end{equation}
 \item imposing that the $\phi_{i\pm}^j$ have appropriate analyticity properties
 at the poles and zeroes of $R(\omega(\tau))$. This yields $8n$ conditions.
 
\end{enumerate} 
 
 \noindent
 Thus, in total, there are $8n + 4$ conditions. 
 They will uniquely determine 
  the value of the $8n + 4$ constants, provided that
 these $8n + 4$ conditions, written as a linear system for the  $8n + 4$ constants, form a linear
 system whose determinant is different from zero.  Then, solving the linear 
systems $S_{1,2}$ results in explicit expressions for $M_{\pm} (\tau)$ in \eqref{mmpm}.
  There may exist points/curves in the $(\rho, v)$ 
 plane where the determinant is zero, in which case there is a breakdown of canonical factorizability at these locations.  Note that since we are dealing with linear systems, we are guaranteed to be able to deduce when ${\cal M}$
 is canonically factorizable, and when not.

In the above, we took $R$ to have only simple zeroes and simple poles. We can easily generalise the
above discussion
to the case when $R$ contains zeroes/poles of higher order.  Using Liouville's theorem, higher order poles can be dealt with
in a similar manner as with simple poles.
 
\vskip 3mm
 
The explicit solution to the systems $S_{1,2}$ determines the explicit form of $M_{\pm}$.  The solution 
to Einstein's field equations is then read off from 
\begin{equation}
M(\rho,v) = M_- (\tau = \infty) \;.
\end{equation}
Let us therefore determine $M_- (\tau = \infty)$. Denoting $[p_j/\tau^{2n}]_{|_{\tau = \infty}} = A_j$ and using $R(\omega(\infty))=1$, we obtain
\begin{eqnarray}
\label{Mrv}
M(\rho, v) &=& 
\begin{pmatrix}
\phi_{1-}^1(\infty) & \quad \phi_{1-}^2(\infty)\\
\phi_{2-}^1 (\infty)& \quad \phi_{2-}^2(\infty)
\end{pmatrix} \\
&=& \tfrac12 \,
\begin{pmatrix}
d_{1-} (\infty)\, A_1  +  d_{2-} (\infty) \, A_2\qquad
& \qquad  d_{1-} (\infty) \, A_3 + d_{2-} (\infty) \, A_4
\\
d_{1-} (\infty)\, A_1 - d_{2-} (\infty) \, A_2 \qquad
& \qquad  d_{1-}(\infty) \, A_3 - d_{2-} (\infty) \, A_4
\end{pmatrix} \;, \nonumber
\end{eqnarray}
where $d_{1-}, d_{2-}$ are evaluated at $\tau = \infty$. 
The constants $A_1, A_2, A_3, A_4$ are not all independent.  The constants $A_3$ and $A_4$ are related
to $A_1$ and $A_2$, as follows.
Recall that $M(\rho, v)$ satisfies $M= M^{\natural} = M^T$, which implies
\begin{equation}
d_{1-} (\infty) \, A_3+ d_{2-} (\infty) \, A_4 = 
d_{1-} (\infty) \, A_1  - d_{2-} (\infty) \, A_2 \;.
\label{relA1}
\end{equation}
In addition, since $\det M =1$ ($M \in SL(2, \mathbb{R})/SO(2)$),  $M$ has the form
\begin{eqnarray}
M =
\begin{pmatrix}
\Delta + B^2/\Delta & \quad B/\Delta\\
B/\Delta  & \quad  \Delta^{-1}
\end{pmatrix}
\;.
\label{Meinst}
\end{eqnarray}
This results in
\begin{equation}
d_{1-} (\infty) \, A_3 - d_{2-} (\infty) \, A_4
= 4  \, \frac{ \left( 1 + \tfrac14 
\left( d_{1-} (\infty) \, A_1 - d_{2-} (\infty) \, A_2\right)^2
\right)}{ d_{1-} (\infty) \, A_1 + d_{2-} (\infty) \, A_2 } \;.
\label{relA2}
\end{equation}
Then, $M$ in \eqref{Mrv} can be expressed in terms of the constants $A_1$ and $A_2$.

\vskip 3mm

Let us return to \eqref{Mmon}:
the functions $a$ and $b$ are continuous
on $\Gamma_{\omega}$, and $a \pm b R$ are non-vanishing on $\Gamma_{\omega}$, as required
for monodromy matrices; the only extra condition imposed in the above was
that $R$ is a rational function of $\omega$, as in the Schwarzschild case. Once $R$ is chosen,
we still have the freedom to pick $a$ and $b$ from a broad class of functions.  Each such choice
will, upon canonical factorization, result in a solution to Einstein's field equations. 
Thus, by changing $a$ and $b$ we obtain a large class of solutions to
the gravitational field equations. Note that
small changes in $a$ and $b$ may result in highly non-trivial changes of the space-time
solution.  We illustrate this in the next subsubsection by picking combinations $a \pm b R$ that possess
an essential singularity at $\omega =0$ and depend on a free parameter $\xi$.
This describes a deformation of the monodromy matrix associated
with the Schwarzschild solution. The canonical factorization of this deformed monodromy matrix
yields a complicated
space-time metric.

\subsubsection{Deforming the monodromy matrix of the Schwarzschild solution}

Let us now consider a concrete example. We take
\begin{eqnarray}
R(\omega) = \frac{\omega +m}{\omega -m} \;\;\;,\;\;\; a(\omega) = \sinh \frac{\xi}{\omega} \;\;\;,\;\;\;
b(\omega) R(\omega) = \cosh \frac{\xi}{\omega} \;,
\label{defschwarz}
\end{eqnarray}
where $m \in \mathbb{R} , \; \xi \in \mathbb{R}$.
When $\xi =0$, the resulting monodromy matrix ${\cal M}(\omega)$ is the one associated to the
Schwarzschild solution \cite{Maison:1988zx}
in the region $v > m >0, \, \rho >0$,
\begin{eqnarray}
{\cal M }_{\rm Schwarzschild} = 
\begin{pmatrix}
R^{-1} & \; 0\\
0  & \; R
\end{pmatrix}
\;.
\end{eqnarray}
When $\xi \neq 0$, the monodromy matrix represents a deformation of ${\cal M }_{\rm Schwarzschild}$, with the property that at $\omega \rightarrow \infty$,  ${\cal M }$ approaches $\mathbb{I}$,
which corresponds to flat space-time. 
As soon as $\xi \neq 0$, the monodromy matrix ceases to be diagonal.
By picking a deformation such that $a$ and $b \, R$ are given in terms of exponentials of rational functions
of $\omega$, we insure that the matrix $D$ can be easily factorized canonically.
The diagonal matrix $D$ has entries
\begin{equation}
d_1 = e^{\frac{\xi}{\omega}} \;\;\;\;,\;\;\;
d_2 = -\frac{1}{d_1} = -e^{-\frac{\xi}{\omega}} \;.
\end{equation}
Using  \eqref{specze}, we obtain
\begin{eqnarray}
d_{1-} = e^{\frac{2\xi \beta}{\rho(\tau - \tau_0^+)}} &\;\;\;,\;\;\;&
d_{1+} = e^{\frac{2\xi \alpha}{\rho(\tau - \tau_0^-)}} \;,
\nonumber\\
d_{2-} = \frac{1}{d_{1-}} =  e^{-\frac{2\xi \beta}{\rho(\tau - \tau_0^+)}}
&\;\;\;,\;\;\;&
d_{2+} = -\frac{1}{d_{1+}} = - e^{-\frac{2\xi \alpha}{\rho(\tau - \tau_0^-)}}\;,
\end{eqnarray}
where we made a choice of signs, and 
where we introduced
\begin{equation}
\alpha = \frac{\tau_0^-}{\tau_0^+ - \tau_0^-} \;\;\;,\;\;\; \beta = - \frac{\tau_0^+}{\tau_0^+ - \tau_0^-} \;.
\end{equation}
The values $\tau_0^{\pm}$ are the two values associated to $\omega =0$ through
the algebraic curve \eqref{omtau},
 \begin{equation}
 \omega = - \frac{\rho}{2} \frac{(\tau - \tau_0^+)(\tau - \tau_0^-)}{\tau} \;.
 \label{specze}
 \end{equation}
They 
are given by (recall that $\rho > 0$)
\begin{equation}
\tau_0^{\pm} = \frac{v \pm \sqrt{v^2 + \rho^2}}{\rho} \;.
\label{valuetau0}
\end{equation}
Here, $\tau_0^+$ denotes the value in the interior of the unit disc in the $\tau$ plane, while
 $\tau_0^-$ denotes the value in the exterior of the unit disc.  The choice of the sign in 
\eqref{valuetau0} is then correlated with the region in the 
 $(\rho,v)$ plane.  
 Similarly, we introduce the two values  $\tau_1^{\pm}$ associated to $\omega =m$,
 \begin{equation}
\tau_1^{\pm} = \frac{(v-m) \pm \sqrt{(v-m)^2 + \rho^2}}{\rho} \;,
\label{valuetau1}
\end{equation}
and the two values  $\tau_2^{\pm}$ associated to $\omega =- m$,
 \begin{equation}
\tau_2^{\pm} = \frac{(v+m) \pm \sqrt{(v+m)^2 + \rho^2}}{\rho} \;.
\label{valuetau2}
\end{equation}

Multiplying the two linear systems $S_{1,2}$ in \eqref{Spm} by $R^{-1}$, we obtain
\begin{eqnarray}
\label{spmsch}
&&S_1: \left\{
\begin{array}{rcl}
 d_{1+}\phi_{2+}^1 + R^{-1} d_{1+}\phi_{1+}^1  = 
(\phi_{1-}^1 + R^{-1} \phi_{2-}^1 )/d_{1-}
 = \frac{A_1\tau^2 + B_1\tau + C_1}{(\tau - \tau_2^+)(\tau - \tau_2^-)} \;, 
 \vspace{2mm} \\
d_{2+}\phi_{2+}^1 - R^{-1} d_{2+}\phi_{1+}^1 = (\phi_{1-}^1 - 
R^{-1} \phi_{2-}^1 )/d_{2-} = \frac{A_2\tau^2 + B_2\tau + C_2}{(\tau - \tau_2^+)(\tau - 
\tau_2^-)} \;,
\end{array}\right. \nonumber\\
&&S_2: \left\{
\begin{array}{rcl}
d_{1+}\phi_{2+}^2 + R^{-1} d_{1+}\phi_{1+}^2 = (\phi_{1-}^2 + R^{-1} \phi_{2-}^2)/d_{1-} = \frac{A_3\tau^2 + B_3\tau + C_3}{(\tau - \tau_2^+)(\tau - \tau_2^-)} \;, 
\vspace{2mm} \\
d_{2+}\phi_{2+}^2 - R^{-1} d_{2+}\phi_{1+}^2 = (\phi_{1-}^2 - R^{-1} \phi_{2-}^2)/d_{2-} = \frac{A_4\tau^2 + B_4\tau + C_4}{(\tau - \tau_2^+)(\tau - 
\tau_2^-)} \;, 
\end{array}\right. 
\end{eqnarray}
which results in
\begin{eqnarray}
&&\phi_{1+}^1 = d_{1+}\frac{A_2\tau^2 + B_2\tau + C_2}{2(\tau - \tau_1^+)(\tau - \tau_1^-)} - d_{2+}\frac{A_1\tau^2 + B_1\tau + C_1}{2(\tau - \tau_1^+)(\tau - \tau_1^-)} \;,\nonumber\\
&& \phi_{1+}^2 = d_{1+}\frac{A_4\tau^2 + B_4\tau + C_4}{2(\tau - \tau_1^+)(\tau - \tau_1^-)} - d_{2+}\frac{A_3\tau^2 + B_3\tau + C_3}{2(\tau - \tau_1^+)(\tau - \tau_1^-)} \;,
\nonumber\\
&&\phi_{2+}^1 = -d_{1+}\frac{A_2\tau^2 + B_2\tau + C_2}{2(\tau - \tau_2^+)(\tau - \tau_2^-)} - d_{2+}\frac{A_1\tau^2 + B_1\tau + C_1}{2(\tau - \tau_2^+)(\tau - \tau_2^-)} \;,
\nonumber\\
&&\phi_{2+}^2 = -d_{1+}\frac{A_4\tau^2 + B_4\tau + C_4}{2(\tau - \tau_2^+)(\tau - \tau_2^-)} - d_{2+}\frac{A_3\tau^2 + B_3\tau + C_3}{2(\tau - \tau_2^+)(\tau - \tau_2^-)} \;,
\nonumber\\
&&\phi_{1-}^1 = \left(\frac{1}{d_{1-}}\right)
\frac{A_2\tau^2 + B_2\tau + C_2}{2(\tau - \tau_2^+)(\tau - \tau_2^-)} + 
\left(\frac{1}{d_{2-}}\right)
\frac{A_1\tau^2 + B_1\tau + C_1}{2(\tau - \tau_2^+)(\tau - \tau_2^-)} \;,
\nonumber\\
&&\phi_{1-}^2 = \left(\frac{1}{d_{1-}} \right)
\frac{A_4\tau^2 + B_4\tau + C_4}{2(\tau - \tau_2^+)(\tau - \tau_2^-)} + 
\left( \frac{1}{d_{2-}} \right) \frac{A_3\tau^2 + B_3\tau + C_3}{2(\tau - \tau_2^+)(\tau - \tau_2^-)} \;,
\nonumber\\
&&\phi_{2-}^1 = - \left( \frac{1}{d_{1-}} \right)
\frac{A_2\tau^2 + B_2\tau + C_2}{2(\tau - \tau_1^+)(\tau - \tau_1^-)} + \left( \frac{1}{d_{2-}} \right) \frac{A_1\tau^2 + B_1\tau + C_1}{2(\tau - \tau_1^+)(\tau - \tau_1^-)} \;, 
\nonumber\\
&&\phi_{2-}^2 = - \left( \frac{1}{d_{1-}} \right) \frac{A_4\tau^2 + B_4\tau + C_4}{2(\tau - \tau_1^+)(\tau - \tau_1^-)} + \left( \frac{1}{d_{2-}} \right) \frac{A_3\tau^2 + B_3\tau + C_3}{2(\tau - \tau_1^+)(\tau - \tau_1^-)} \;.
\label{phiABC}
\end{eqnarray}
The $12$ constants $A_j, B_j, C_j, \; j=1,2,3,4$ are determined by imposing the conditions described below \eqref{Spm}.
Imposing the normalization condition $M_+ (\tau =0) = \mathbb{I}$ yields
\begin{equation}
C_1 = - d_{1+} (\tau =0) \;\;\;,\;\;\; C_2 = \frac{1}{C_1} \;\;\;,\;\;\; C_3 = - d_{1+} (\tau =0) \;\;\;,\;\;\; C_4 = - \frac{1}{C_3} \;.
\end{equation}
Imposing that the $\phi^j_{i \pm}$ in \eqref{phiABC} have the appropriate analyticity requirements
at $\tau_{1,2}^{\pm}$
determines the coefficients $A_j, B_j$. 
We find that the $A_j$ and $B_j$ take the form
$A_j = \tilde{A}_j/ K, \; B_j = \tilde{B}_j/K \; ( j=1,2, 3, 4)$, where
\begin{eqnarray}
K &=& \tau_1^+ \, \tau_2^+ \Big\{
2 C_1^2 D_{1+} D_{2+} \, \left(1 + (\tau_1^+)^2\right)  \left(1 + (\tau_2^+)^2 \right) \nonumber\\
&& \qquad \quad + C_1^2
\left(-C_1^2 (\tau_1^+ - \tau_2^+)^2 + D_{2+}^2 \left(1 + \tau_1^+ \tau_2^+ \right)^2 \right)
\nonumber\\
&& \qquad \quad  + D_{1+}^2 \left( - D_{2 +}^2 (\tau_1^+ - \tau_2^+)^2 + C_1^2  \left(1 + \tau_1^+ \tau_2^+ \right)^2 \right)
\Big\} \;,
\label{expAi}
\end{eqnarray}
and similar expressions for the ${\tilde A}_j$ and for the ${\tilde B}_j$.  Here,
we introduced the notation
\begin{equation}
D_{1+} = d_{1+}^2(\tau = \tau_1^+) \;\;,\;\; D_{2+} = d_{1+}^2(\tau = \tau_2^+) \;\;,\;\;
D_{1-} = d_{1-}^2(\tau = \tau_1^-) \;\;,\;\; D_{2-} = d_{1-}^2(\tau = \tau_2^-) \;,
\end{equation}
for convenience.
Thus, as long as $K$ does not vanish,
we have obtained a canonical factorization.

We recall that $A_3$ and $A_4$ have to satisfy the relations \eqref{relA1} and \eqref{relA2}.
We proceed to verify that $A_1, A_2, A_3, A_4$ indeed satisfy these relations.
In doing so, we use the relations
\begin{equation} 
\frac{d_{1+}(\tau_1^+)}{d_{1-}(\tau_1^-)} = d_{1+} (\tau =0) = \frac{d_{1+}(\tau_2^+)}{d_{1-}(\tau_2^-)} \;.
\end{equation}

Next, we turn to $M(\rho, v)$. 
Using $d_{1-} (\infty) = d_{2-} (\infty) =1$ we obtain from \eqref{Mrv},
\begin{eqnarray}
M(\rho, v)  = \tfrac12 
\begin{pmatrix}
A_1 + A_2\qquad
& \qquad  A_1 - A_2
\\
A_1 - A_2 \qquad & \qquad  
 4 \, \frac{ \left( 1 + \tfrac14 
\left( A_1 - A_2 \right)^2
\right)}{ \left( A_1 + A_2  \right)}
\end{pmatrix} \;.
\label{Mdf}
\end{eqnarray}
The associated space-time solution is described by a stationary line element of the form
\begin{eqnarray}
ds^2_4 = - \Delta \left(dt + \sigma \right)^2 + \Delta^{-1} \, 
\left( e^{\psi} (d\rho^2 + dv^2) + \rho^2 \, d\phi^2 \right) \;,
\label{deformedsch}
\end{eqnarray}
where both $\Delta$ and the one-form $\sigma = \sigma_{\phi} \, d \phi$  are determined in terms of the entries of $M(\rho,v)$.
Namely, $\Delta$ is read off from \eqref{Mdf} using 
\eqref{Meinst}, while the two-form $F = d \sigma$ is determined by
\begin{equation}
\Delta^2 \, *  F = d B \;,
\label{dualB}
\end{equation}
where $*$ denotes the Hodge dual in three dimensions with respect to the metric \eqref{met3d}.
Finally, $\psi$ is determined by \eqref{psi2}.

The coefficients $A_1$ and $A_2$, and hence $\Delta$ and $B$, are very complicated functions of the Weyl coordinates
$(\rho, v)$.  To first order in the deformation parameter $\xi$ we obtain,
\begin{eqnarray}
\Delta &=& \frac{\tau_2^+}{\tau_1^+} + {\cal O} (\xi^2) \;, \nonumber\\
B &=&2 \xi \, \frac{  \tau_2^+}{\rho( \tau_0^+ - \tau_0^-)   \, \tau_1^+
 (1 + \tau_1^+ \,  \tau_2^+) (\tau_1^+ - \tau_0^-) (\tau_2^+ - \tau_0^-) } \nonumber\\
&& \times \Big(  \tau_2^+ \,  \tau_0^-  - (\tau_0^-)^2 + \tau_1^+ (\tau_2^+ + \tau_0^-) 
(1 + \tau_2^+ \tau_0^-) -
 (\tau_2^+)^2 (1 + (\tau_0^-)^2)  \nonumber\\
&& \qquad  -  
   (\tau_1^+)^2 (1 + (\tau_2^+)^2  
      - (\tau_2^+) \, \tau_0^-   + (\tau_0^-)^2) \Big)  + {\cal O} (\xi^2)  \;,
   \end{eqnarray}
while $\psi$ is undeformed at first order in $\xi$. Since $B$ is non-vanishing at first order in $\xi$, it induces
a non-vanishing two-form $F$ given by \eqref{dualB}.  We have verified that $F$ is closed,
i.e. $d F = 0$, so that locally, $F = d \sigma$, with $\sigma = \sigma_{\phi} \, d \phi$.

The above shows that the Riemann-Hilbert factorization method constitutes a feasible and explicit method for obtaining
deformed solutions to the field equations of gravitational field theories that may be hard to obtain by direct means,
i.e. by directly solving the non-linear field equations.

We conclude by 
verifying the validity of the substitution rule. To this end, we focus on the region $v > m >0$, and consider the limit
$\rho \rightarrow 0^+$. We obtain
\begin{eqnarray}
\tau_{1}^+ &\rightarrow& - \tfrac12 \, \frac{\rho}{v-m} \rightarrow 0\;\;\;,\;\;\; 
\tau_{2}^+ \rightarrow - \tfrac12 \, \frac{\rho}{v+m} \rightarrow 0 \;, \nonumber\\
d_{1+}(0) &\rightarrow& e^{\xi/v} \;\;\;,\;\;\; d_{1+} (\tau_1^+) \rightarrow d_{1+}(0) \;\;\;,\;\;\;
d_{1+} (\tau_2^+) \rightarrow d_{1+}(0) \;.
\end{eqnarray}
Using this, we find in the limit $\rho \rightarrow 0^+$,
\begin{eqnarray}
{\tilde A}_1 &=& - 2 C_1^7 \, \tau_2^+ \left(\tau_1^+ + \tau_2^+ \right) + 2 C_1^5 \, \tau_2^+ 
\left(\tau_1^+ - \tau_2^+ \right) \;, \nonumber\\
{\tilde A}_2 &=&  2 C_1^7 \, \tau_2^+ \left(\tau_1^+ - \tau_2^+ \right) - 2 C_1^5 \, \tau_2^+ 
\left(\tau_1^+ + \tau_2^+ \right) \;, \nonumber\\
K & \rightarrow & 4 C_1^6 \, \tau_1^+ \tau_2^+ \;,
\end{eqnarray}
and hence
\begin{eqnarray}
\tfrac12 (A_1 + A_2) \rightarrow \left(\frac{v-m}{v+m} \right) \, \cosh\frac{\xi}{v} \;, \nonumber\\
\tfrac12 (A_1 - A_2) \rightarrow \sinh\frac{\xi}{v} \;.
\end{eqnarray}
Thus we infer
\begin{eqnarray}
\lim_{\rho \rightarrow 0^+} M(\rho, v) = {\cal M}(v) \;,
\end{eqnarray}
which verifies the validity of the substitution rule in the region $v > m >0$.

\subsection*{Acknowledgements}

\noindent
We would like to thank Cristina C\^amara for valuable suggestions and discussions, and Thomas Mohaupt and Suresh Nampuri
for valuable comments.  J.C.S. gratefully acknowledges the support of the Gulbenkian
Foundation through the scholarship program
{\tt Novos Talentos em Matem\'atica}. 
This work was partially supported by FCT/Portugal through UID/MAT/04459/2013.

\appendix

\section{Explicit factorization \label{explfact}}

We perform the explicit factorization of \eqref{matrixtilH} by 
solving the associated vectorial factorization problem \eqref{vecfact} column by column,
and using (a generalized version of) Liouville's theorem (see
 the lemma on page 14 of \cite{Camara:2017hez}). We impose the
normalization condition ${\tilde M}_+ (\tau =0) = \mathbb{I}$.

To this end, we introduce the 
following notation: we denote by $\phi_{i\pm}^j$  the element in line $i$ and column $j$ of the matrix ${\tilde M}_+^{-1}$ or ${\tilde M}_-$ in \eqref{matrixtilH}. We use
\begin{eqnarray}
H_1 (\omega(\tau)) = h_1 \frac{(\tau - \tau_{\tilde Q}^-)(\tau - \tau_{\tilde Q}^+)}{(\tau - \tau_0^-)(\tau - \tau_0^+)} \;,\nonumber\\
H_2 (\omega(\tau)) = h_2 \frac{(\tau - \tau_{\tilde P}^-)(\tau - \tau_{\tilde P}^+)}{(\tau - \tau_0^-)(\tau - \tau_0^+)}\;.
\end{eqnarray}

The third line of \eqref{matrixtilH} yields
\begin{eqnarray}
\phi_{1+}^1 &=& 1 \;\;\;,\;\;\;  \phi_{3-}^1 = -1 \;, \nonumber\\
\phi_{1+}^2 &=& 0 \;\;\;,\;\;\; \phi_{3-}^2 = 0 \;, \nonumber\\
\phi_{1+}^3 &=& 0 \;\;\;,\;\;\;  \phi_{3-}^3 = 0 \;.
\end{eqnarray}


The second line of \eqref{matrixtilH} yields the following.
First we consider
\begin{eqnarray}
\phi_{2-}^2 = - \frac{H_1}{H_2} \, \phi_{2+}^2 = \frac{\alpha_1 \, \tau + \beta_1}{\tau - \tau_{\tilde P}^+} \;,
\end{eqnarray}
by Liouville's theorem.  This is solved by
\begin{eqnarray}
\phi_{2+}^2 &=& \frac{\tau_{\tilde Q}^-}{\tau_{\tilde P}^-}
\left( \frac{\tau - \tau_{\tilde P}^-}{\tau - \tau_{\tilde Q}^-} \right) \;, \nonumber\\
\phi_{2-}^2 &=& -\frac{h_1}{h_2}  \, \frac{\tau_{\tilde Q}^-}{\tau_{\tilde P}^-}
\left(
\frac{\tau - \tau_{\tilde Q}^+}{\tau - \tau_{\tilde P}^+} \right) \;.
\end{eqnarray}
Next, we consider
\begin{equation}
\phi_{2-}^1 = - \sqrt{2} \, H_1 \, \phi_{1+}^1 - \frac{H_1}{H_2} \,  \phi_{2+}^1 =
\frac{\alpha_2 \, \tau^2 + \beta_2 \, \tau + \gamma_2}{(\tau - \tau_{\tilde{P}^+} ) 
(\tau - \tau_0^+) }  \;,
\end{equation}
by Liouville's theorem.  This is solved by
\begin{eqnarray}
\phi_{2+}^1 &=& -
\sqrt{2} \, h_2 \,
\left( \frac{\tau - \tau_{\tilde P}^-}{\tau - \tau_{\tilde Q}^-} \right)
\, \frac{\tau}{\tau - \tau_0^-} \, 
\frac{(\tau_0^+ - \tau_{\tilde Q}^+) (\tau_0^- - \tau_{\tilde P}^+)}{
\tau_{\tilde Q}^+ \, (\tau_0^+ - \tau_0^-)} \;, \nonumber\\
\phi_{2-}^1 &=& - \frac{\sqrt{2} \, h_1 }{ \tau_{\tilde Q}^+}
\left( \frac{\tau - \tau_{\tilde Q}^+}{\tau - \tau_{\tilde P}^+} \right)
 \frac{({\tilde \alpha} \, \tau - \tau_{\tilde P}^+ \tau_0^+  )
}{(\tau - \tau_0^+)} \;, \nonumber\\
\end{eqnarray}
where
\begin{equation}
{\tilde \alpha} = \frac{ \tau_{\tilde P}^+ \, ( \tau_0^+ - \tau_{\tilde Q}^+) + \tau_{\tilde Q}^+ \tau_0^+
+ 1}{\tau_0^+ - \tau_0^-} \;.
\end{equation}
In addition, we find
\begin{eqnarray}
\phi_{2+}^3 = 0 \;\;\;,\;\;\;
\phi_{2-}^3 = 0 \;.
\end{eqnarray}

Next, we turn to the first line of \eqref{matrixtilH}. 
We find
\begin{equation}
\phi_{3+}^3 = 1 \;\;\;,\;\;\;
\phi_{1-}^3 = -1 \;.
\end{equation}
We also get
\begin{equation}
\sqrt{2} \, H_1 \, \phi_{2 +}^2 - \phi_{3 +}^2 = \phi_{1-}^2 = \frac{\alpha_3 \, \tau + \beta_3}{\tau -
\tau_0^+} \;,
\end{equation}
by Liouville's theorem.  This is solved by
\begin{eqnarray}
\phi_{3+}^2 &=& \sqrt{2} h_1
 \, \frac{\left( \tau_{\tilde Q}^- \, ( \tau_{\tilde P}^- -  \tau_0^-)  -
\tau_0^+ \tau_{\tilde P}^- - 1 \right)}{ \tau_{\tilde P}^- (\tau_0^+ -  \tau_0^-) } \, \frac{\tau}{
( \tau -  \tau_0^-) } \;, \nonumber\\
\phi_{1-}^2 &=& \sqrt 2 h_1 \, 
 \, \frac{ \left[ \tau \left( \tau_0^+ \, ( \tau_{\tilde P}^- +  \tau_{\tilde Q}^-)  -
\tau_{\tilde Q}^-  \tau_{\tilde P}^- + 1 
\right) \right] - \tau_0^+ \tau_{\tilde P}^- (\tau_0^+ - \tau_ 0^-)
}{ \tau_{\tilde P}^- (\tau_0^+ -  \tau_0^-)  ( \tau -  \tau_0^+)
} \;.
\end{eqnarray}
Finally, we also obtain
\begin{equation}
H_1 H_2 \phi_{1+}^1 + \sqrt{2} H_1 \, \phi_{2 +}^1 - \phi_{3 +}^1 = \phi_{1-}^1 = 
\frac{\alpha_4 \tau^2 + \beta_4 \tau + \gamma_4}{(\tau - \tau_0^+)^2} \;,
\end{equation}
by Liouville's theorem.  This yields
\begin{eqnarray}
\phi_{1-}^1 &=& 
\frac{\alpha_4 \tau^2 + \beta_4 \tau + \gamma_4}{(\tau - \tau_0^+)^2} 
\end{eqnarray}
as well as 
\begin{eqnarray}
\phi_{3+}^1 &=& \frac{1}{( \tau - \tau_0^+)^2 (\tau - \tau_0^-)^2}
\big[ - (\tau - \tau_0^-)^2 (\alpha_4 \tau^2 + \beta_4 \tau + \gamma_4)  \\
&& \quad   \qquad \qquad \qquad \qquad + h_1 h_2 (\tau - \tau_{\tilde Q}^-)  (\tau - \tau_{\tilde P}^-)
(\tau - \tau_{\tilde Q}^+)  (\tau - \tau_{\tilde P}^+) \nonumber\\
&&  \qquad \qquad \qquad \quad  \qquad -2 h_1 h_2 \frac{(\tau_0^+ - \tau_{\tilde Q}^+)(\tau_0^- - \tau_{\tilde P}^+)}{
(\tau_0^+ - \tau_0^-) \tau_{\tilde Q}^+} \, \tau \, (\tau - \tau_0^+) (\tau -  \tau_{\tilde Q}^+ )
(\tau - \tau_{\tilde P}^-)
\big] \;, \nonumber
\end{eqnarray}
where the numerator of $\phi_{3+}^1 $ has to have a double zero at $\tau = \tau_0^+$,
which results in 
\begin{eqnarray}
\gamma_4 &=& h_1 h_2 \, (\tau_0^+)^2 \;, \nonumber\\
\beta_4 \, \tau_0^+ &=& - \alpha_4 (\tau_0^+)^2 - \gamma_4 + \frac{h_1 h_2}{(\tau_0^+ - 
\tau_0^-)^2} (\tau_0^+ - \tau_{\tilde Q}^-) (\tau_0^+ - \tau_{\tilde P}^-) 
(\tau_0^+ - 
\tau_{\tilde Q}^+) (\tau_0^+ - \tau_{\tilde P}^+) \;, \nonumber\\
\alpha_4 &=& h_1 h_2 \left(1 - \frac{2 {\tilde Q}}{h_1 \, \rho \, (\tau_0^+ - \tau_0^-)} \right)
\left(1 - \frac{2 {\tilde P}}{h_2 \, \rho \, (\tau_0^+ - \tau_0^-)} \right) 
\nonumber\\
&& 
- 2 h_1 h_2  \, \frac{(\tau_{\tilde Q}^+ - \tau_{\tilde P}^+ ) (\tau_0^+ - \tau_{\tilde P}^-)
(\tau_0^+ - \tau_{\tilde Q}^+)}{\tau_{\tilde Q}^+ (\tau_0^+ - \tau_0^- )^2} \;.
\end{eqnarray}

Having obtained the factorization of \eqref{matrixtilH}, we extract 
the matrix $M(\rho,v)$ given in \eqref{matr}. In obtaining the expressions given in \eqref{m123},
we
used the following relation,
\begin{eqnarray}
\frac{2 \tilde Q}{\rho } &=& \frac{(\tau_0^+ - \tau_{\tilde Q}^+)(\tau_0^- - \tau_{\tilde Q}^+)}{\tau_{\tilde Q}^+}
= - \frac{(\tau_0^+ - \tau_{\tilde Q}^-)(\tau_0^+ -  \tau_{\tilde Q}^+)}{\tau_0^+} 
= \tau_{\tilde Q}^+ + \tau_{\tilde Q}^- - \tau_0^+ - \tau_0^- \;.
\end{eqnarray}

\providecommand{\href}[2]{#2}\begingroup\raggedright\endgroup

\end{document}